\documentclass[prd,preprint,superscriptaddress,preprintnumbers,eqsecnum,showpacs,nofootinbib,nobibnotes]{revtex4-1}
\usepackage{amsfonts,amsmath,amssymb,bm,natbib}
\usepackage{graphicx}
\usepackage{slashed} 
\usepackage{subfigure}
\usepackage{color}

\newcommand{\be}{\begin{equation}}
\newcommand{\bea}{\begin{eqnarray}}
\newcommand{\ee}{\end{equation}}
\newcommand{\eea}{\end{eqnarray}}

\def\s#1{{\scriptscriptstyle #1}}

\def\1eq#1{Eq.(\ref{#1})}

\def\2eqs#1#2{Eqs.~(\ref{#1}) and~(\ref{#2})}
\def\3eqs#1#2#3{Eqs.~(\ref{#1}),~(\ref{#2}) and~(\ref{#3})}

\def\ie{{\it i.e.}, }
\def\eg{{\it e.g.}, }

\def\n#1{({\it #1}\,)}

\newcommand{\STI}{\textnormal{\tiny \textsc{STI}}}
\newcommand{\Tr}{\textnormal{\tiny \textsc{T}}}
\newcommand{\FBC}{\rm{\s{FBC}}}

\begin{document}

\title{Quark gap equation with non-abelian Ball-Chiu vertex}

\author{A.~C. Aguilar}
\affiliation{University of Campinas - UNICAMP, 
Institute of Physics ``Gleb Wataghin'',
13083-859 Campinas, SP, Brazil}

\author{J. C. Cardona}
\affiliation{University of Campinas - UNICAMP, 
Institute of Physics ``Gleb Wataghin'',
13083-859 Campinas, SP, Brazil}

\author{M. N. Ferreira}
\affiliation{University of Campinas - UNICAMP, 
Institute of Physics ``Gleb Wataghin'',
13083-859 Campinas, SP, Brazil}

\author{J. Papavassiliou}
\affiliation{\mbox{Department of Theoretical Physics and IFIC, 
University of Valencia and CSIC},
E-46100, Valencia, Spain}

\begin{abstract}

  The  full  quark-gluon  vertex  is  a  crucial  ingredient  for  the
  dynamical generation of  a constituent quark mass  from the standard
  quark gap  equation, and its  non-transverse part may  be determined
  exactly  from   the  nonlinear   Slavnov-Taylor  identity   that  it
  satisfies.   The resulting  expression involves  not only  the quark
  propagator, but also the ghost dressing function and the quark-ghost
  kernel, and  constitutes the non-abelian extension  of the so-called
  ``Ball-Chiu vertex'', known from QED.   In the present work we carry
  out  a detailed  study  of the  impact  of this  vertex  on the  gap
  equation and the quark masses  generated from it, putting particular
  emphasis on the contributions directly related with the ghost sector
  of  the   theory,  and   especially  the  quark-ghost   kernel.   In
  particular, we set up and solve  the coupled system of six equations
  that determine  the four form factors  of the latter kernel  and the
  two typical Dirac structures composing  the quark propagator. Due to
  the    incomplete     implementation    of     the    multiplicative
  renormalizability  at the  level of  the gap  equation, the  correct
  anomalous  dimension of  the  quark mass  is  recovered through  the
  inclusion  of a  certain  function, whose  ultraviolet behavior  is
  fixed,  but its  infrared  completion is  unknown; three  particular
  Ans\"atze for this function are  considered, and their effect on the
  quark  mass and  the  pion  decay constant  is  explored.  The  main
  results  of this  study indicate  that the  numerical impact  of the
  quark-ghost kernel is considerable; the transition from a tree-level
  kernel to  the one  computed here  leads to a  20\% increase  in the
  value of the quark mass  at the origin.  Particularly interesting is
  the  contribution  of  the  fourth  Ball-Chiu  form  factor,  which,
  contrary to the abelian case, is non-vanishing, and accounts for 10\%
  of the total constituent quark mass.

\end{abstract}


\maketitle

\section{\label{sec:intro} Introduction}

The dynamical breaking of chiral  symmetry and the generation of a
constituent mass for the  quarks represent  two of  the most
important emergent  phenomena in  QCD, and the  detailed study  of the
nonperturbative dynamics associated with them has been the focal point
of  countless  articles spanning several decades~\cite{Nambu:1961tp,Lane:1974he,Politzer:1976tv,Miransky:1981rt,Atkinson:1988mw,Brown:1988bm,Williams:1989tv,Papavassiliou:1991hx,Hawes:1993ef,Roberts:1994dr,Natale:1996eu,Fischer:2003rp,Maris:2003vk, Aguilar:2005sb,Bowman:2005vx,Sauli:2006ba,Cornwall:2008da,Aguilar:2010cn,Cloet:2013jya,Mitter:2014wpa,Braun:2014ata,Heupel:2014ina,Binosi:2016wcx}.  
One  of  the
standard frameworks employed in this  pursuit is the so-called ``quark
gap-equation'',  namely   the  Schwinger-Dyson  equation   (SDE)~\cite{Roberts:1994dr,Maris:2003vk} that
controls the  evolution of the  quark propagator $S(p)$.  This special
integral equation  is particularly  sensitive to the  ingredients that
compose  its  kernel,  and  in   particular  on  the  details  of  the
fully-dressed   quark-gluon  vertex $\Gamma_{\mu}(q,p_2,-p_1)$
~\cite{Roberts:1994dr}.   This 
latter three-point function  is  built  out  of  twelve  linearly  
independent  tensorial structures~\cite{Ball:1980ay,Kizilersu:1995iz,Davydychev:2000rt,Bermudez:2017bpx}, and  the determination of the  
nonperturbative behavior of
the corresponding  form-factors represents  a major challenge  for the 
contemporary field-theoretic formalisms, both continuous and discrete~\cite{Bender:1996bb,Bhagwat:2004kj,LlanesEstrada:2004jz,Holl:2004qn,Matevosyan:2006bk,Alkofer:2008tt,Aguilar:2010cn,
Hopfer:2013np,Aguilar:2013ac,Rojas:2013tza,Williams:2014iea,Williams:2015cvx,Sanchis-Alepuz:2015qra,Binosi:2016wcx,Aguilar:2016lbe,Fischer:2006ub,Skullerud:2002sk,Skullerud:2002ge,Skullerud:2003qu,Skullerud:2004gp,Lin:2005zd,Kizilersu:2006et,Oliveira:2016muq,Sternbeck:2017ntv}.

The quark-gluon vertex $\Gamma_{\mu}$ satisfies a non-linear Slavnov-Taylor identity (STI),
given by
\mbox{$q^{\mu}\Gamma_{\mu}(q,p_2,\!-p_1\!)\! = \!F(q)[S^{-1}\!(p_1) H\!(q,p_2,\!-p_1\!)\! -\!{\overline H}(\!-q,p_1,\!-p_2\!)S^{-1}\!(p_2)]$}, where $F(q)$ is the dressing function of the
ghost propagator, and $H$ is the so-called quark-ghost kernel,
which consists of four linearly independent tensorial structures, and 
$S^{-1}(p) = A(p)\slashed{p} - B(p)$.
When the ghost sector is switched off (\ie $F=H=1$), the above
STI reduces to the standard Ward-Takahashi identity of QED.
It is common practice to decompose 
$\Gamma_{\mu}$ into two parts, \mbox{$\Gamma_{\mu} = \Gamma^{\STI}_{\mu} + \Gamma^{\Tr}_{\mu}$}, where 
$\Gamma^{\STI}_{\mu}$ {\it saturates} the above STI, while $\Gamma^{\Tr}_{\mu}$ denotes 
the transverse (automatically conserved) part, (\ie $q^{\mu}\Gamma^{\Tr}_{\mu}=0$).
Then, it turns out that the four form factors comprising 
$\Gamma^{\STI}_{\mu}$, to be denoted by $L_i$,
may be expressed entirely in terms of combinations involving $A$, $B$, 
and the form factors of $H$. The $\Gamma^{\STI}_{\mu}$ obtained from the
abelianized version of the STI (setting $F=H=1$) is known in the literature as the
{\it``Ball-Chiu'' vertex}~\cite{Ball:1980ay}, and will be denoted by $\Gamma^{\rm \s{BC}}_{\mu}$.  
In order to establish a clear distinction
between $\Gamma^{\rm \s{BC}}_{\mu}$ and 
the full $\Gamma^{\STI}_{\mu}$, which includes, at least in principle, all ghost related contributions
(and, in particular, those from $H$), 
we will  denominate the latter as the {\it ``non-abelian Ball-Chiu  vertex''}\footnote{In what
follows we will use the terms ``$\Gamma^{\STI}_{\mu}$'' and ``non-abelian Ball-Chiu  vertex''
interchangeably.}.

Since the form factors of $H$, to be denoted by $X_i$, 
constitute an indispensable ingredient for the complete determination 
of $\Gamma^{\STI}_{\mu}$, in a recent work~\cite{Aguilar:2016lbe} a SDE-based procedure was developed  
for their dynamical determination. Specifically, the skeleton expansion of $H$ 
was truncated at its ``one-loop-dressed'' level, and the four $X_i$ were determined
by means of appropriate projections, for arbitrary values of Euclidean momenta.
As a result, one obtained approximate expressions for the form factors of $\Gamma^{\rm \s{STI}}_{\mu}$,
which receive nontrivial contributions from the kernel $H$, whose numerical
impact is quite considerable. In particular, not only is the difference between 
$\Gamma^{\STI}_{\mu}$ and
$\Gamma^{\rm \s{BC}}_{\mu}$ particularly pronounced, but a considerable difference is found also between
$\Gamma^{\STI}_{\mu}$ and the ``minimally non-abelianized'' Ball-Chiu  vertex,
obtained by multiplying $\Gamma^{\rm \s{BC}}_{\mu}$ by $F(q)$; we denote this latter vertex
by $\Gamma^{\FBC}_{\mu} = F(q) \Gamma^{\rm \s{BC}}_{\mu}$~\cite{Fischer:2003rp,Aguilar:2016lbe}.
Note that the resulting
form factors of $\Gamma^{\STI}_{\mu}(q,p_2,-p_1)$ display a completely nontrivial dependence on 
three kinematic variables, chosen to be the moduli of two of the incoming momenta, $p_1$ and $p_2$, and the angle $\theta$ between them. 

Given that $\Gamma_{\mu}$ is known to be particularly relevant for the studies of the phenomena
controlled by the gap equation, it is natural to explore the impact that the $\Gamma^{\rm \s{STI}}_{\mu}$
constructed in~\cite{Aguilar:2016lbe} might have on dynamical chiral symmetry breaking and quark mass generation.
The purpose of the present work is to carry out a detailed quantitative study of this particular question,
adding, at the same time, an extra layer of technical complexity to the considerations presented so far.
Specifically, in the analysis of~\cite{Aguilar:2014lha,Aguilar:2016lbe},  
$S(p)$ was essentially treated as an ``external'' quantity: the corresponding $A$ and $B$ used for the evaluation of
the $X_i$ were obtained from solving a gap equation containing a simplified version of  $\Gamma^{\STI}_{\mu}$.
It is clear, however, that the self-consistent treatment of this problem requires
the solution of a {\it coupled system} of several dynamical equations, given that $S(p)$ enters in the integrals that
determine the form factors of $H$, which, in turn, enter through $\Gamma^{\STI}_{\mu}$ 
in the gap equation that determines $S(p)$. Therefore, in the analysis presented here, we will
consider the intertwined dynamics produced by a system involving six coupled equations
(four determining the $X_i$, and two the $A$ and $B$). 

There are two important issues related to our analysis that need to be emphasized at this point.
First, the gap equation is studied in the {\it chiral limit}, \ie no ``current'' mass, $m_0$, is added to the corresponding equations (see, for example, \1eq{gap}).
Second, the external ingredients used (see subsection~\ref{Num_inputs}) are obtained from ``quenched'' lattice simulations; this simplification
affects both the gluon propagator and the
ghost dressing function, and, indirectly, the form factors of the quark-gluon vertex, and, eventually, the gap equation itself.
Unquenching effects have been taken into account in the context of other
approaches~\cite{Bhagwat:2004kj,Fischer:2005wx,Fischer:2005en,Williams:2015cvx,Cyrol:2017ewj}, and   
can also be treated within our formalism, along the lines presented 
in~\cite{Aguilar:2012rz}. Such a study, however,
lies beyond the main scope of the present work, which focuses on the impact that the fully non-abelianized Ball-Chiu vertex has on the gap equation.

The main findings of our study may be summarized as follows:

\begin{enumerate}

\item The  dynamical quark masses, ${\mathcal M}(p)$,  generated with $\Gamma^{\STI}_{\mu}$ are
always higher than those obtained with the $\Gamma^{\rm \s{BC}}_{\mu}$. 
The precise amount depends on the specific value of $\alpha_s$ employed in the
numerical calculation, but, on the  average, 
the impact of $H$ on ${\mathcal M}(0)$
is of the order of $20\%$ for the cases where  ${\mathcal M}(p)$ is around \mbox{$300-350$ MeV} [see Fig.~\ref{mass_1G}]. 
The quark wave functions follow a similar pattern, with $A^{-1}(p)$  always larger than $A^{-1}_{\FBC}(p)$
[Fig.~\ref{mass_1G}].

\item The results for the vertex form factors, $L_i$, obtained after solving  the coupled system,
display the same qualitative and quantitative behavior found in Ref.~\cite{Aguilar:2016lbe},  
where  $A(p)$ and $B(p)$ were treated as external ingredients [see Fig.~\ref{fig:LSDE}].

\item  The form factor $L_1$ 
  is responsible for generating more than half of the value of  ${\mathcal M}(0)$ ($54\%$),
  whereas $L_2$, and $L_3$ provide  $13\%$ and $23\%$ of the quark mass value at zero momentum, respectively.
Particularly interesting is the considerable contribution originating 
from the inclusion of $L_4$, which is  commonly neglected in the quark SDE studies, 
accounting for $10\%$ of  ${\mathcal M}(0)$.

\item The pion constant decay, $f_{\pi}$, was used as a simple indicator
  of the impact that the inclusion of  $H$ in the construction of  $\Gamma^{\STI}_{\mu}$ might have
  on physical quantities. 
  Our study reveals that the final impact of $H$ is to increase  by \mbox{$10\%$}
of the value of $f_{\pi}$ [see Table~\ref{table_fpi}].

\item All quark masses obtained may be fitted accurately by two very simple formulas, given by Eqs.~\eqref{fit} 
and~\eqref{fit2}, which, at large momenta,
reproduce the well-known power-law behavior~expressed in Eq.~\eqref{fituv} [see Fig.~\ref{figure:fit}].

\end{enumerate}

The article is organized as follows. In section~\ref{sec:background}
we introduce the notation and set up the theoretical framework of this work,  
and review the general structure of the gap equation, together with 
the SDEs for the four form factors, $X_i$.
In section~\ref{Num} we  present  
the numerical treatment of the system of six coupled integral equations, formed by $A(p)$, $B(p)$ and the four $X_i$.
Finally, in section~\ref{sec:c} we present our conclusions.

\section{\label{sec:background} Theoretical ingredients and derivation of the system}

\begin{figure}[!h]
\includegraphics[scale=0.7]{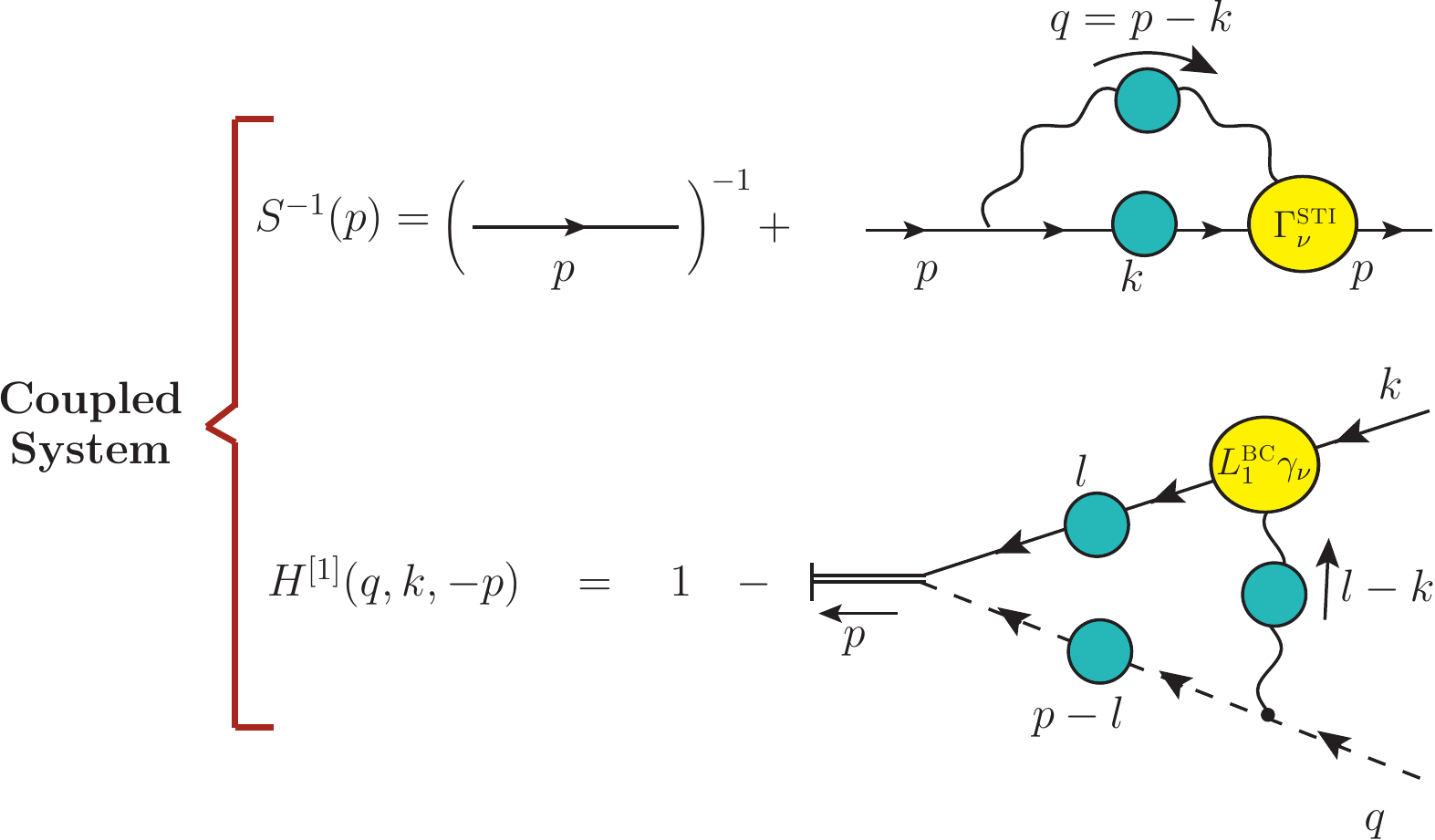}
\vspace{-0.25cm}
\caption{\label{fig:system} The SDE for the quark propagator, $S(p)$ (top), and 
the quark-ghost scattering kernel at one-loop dressed approximation, \mbox{$H^{[1]}(q,k,-p)$} (bottom). The quark-gluon vertex, $\Gamma^{\STI}_{\mu}$, couples  $S(p)$ to \mbox{$H^{[1]}(q,k,-p)$}.}
\end{figure}

In this section we review all ingredients and concepts necessary for arriving at the
system of integral equations that is diagrammatically depicted in Fig.~\ref{fig:system}.

\subsection{\label{sec:gap} Gap equation and quark-gluon vertex }

The full quark propagator can be written as 
\be
S^{-1}(p) =  A(p)\slashed{p} - B(p) \mathbb{I} 
= A(p)[\slashed{p}-{\mathcal{M}}(p) \mathbb{I}] \,,
\label{qAB}
\ee
where $A(p)$ and  $B(p)$ are scalar functions whose ratio defines the
dynamical quark mass function \mbox{${\mathcal{M}}(p)= B(p)/A(p)$}.

The momentum-dependence of $S(p)$, or, equivalently, of the functions $A(p)$ and  $B(p)$,
may be obtained from the quark gap equation, which, in its
{\it renormalized} form, is given by  
\begin{equation}
S^{-1}(p)= Z_{\rm{\s F}}  \slashed{p}  -Z_{1}C_{\rm{\s F}}g^2\!\!\int_k\,
\gamma_{\mu}S(k)\Gamma_{\nu}(q,k,-p)\Delta^{\mu\nu}(q) \,,
\label{gap}
\end{equation}%
where $C_{\rm {\s F}}$ is the Casimir eigenvalue for the fundamental representation, and  
we  have introduced the compact notation for the integral measure  
\be
\int_{k}\equiv\frac{\mu^{\epsilon}}{(2\pi)^{d}}\!\int\!\mathrm{d}^d k,
\label{dqd}
\ee
with $\mu$ the 't Hooft mass, and  $d=4-\epsilon$ the space-time dimension. 
In addition, 
$\Gamma_{\nu}(q,k,-p)$ is the full quark-gluon vertex,
while $Z_1(\mu)$ and $Z_{\rm{\s F}}(\mu)$ are the vertex and the quark wave-function renormalization constants, respectively, and $\mu$ is the renormalization point. Moreover, in the Landau gauge, the full gluon propagator
$\Delta_{\mu\nu}(q)$ is  given by
\begin{align}
i\Delta_{\mu\nu}(q)=-iP_{\mu\nu}(q)\Delta(q); &\quad P_{\mu\nu}(q)=g_{\mu\nu}-\frac{q_\mu q_\nu}{q^2} \,.
\label{propagators}
\end{align}
Note finally that Eq.~\eqref{gap} is expressed in the chiral limit, since it contains no ``current'' quark mass ($m_0=0$). 

$\Gamma^a_{\mu}(q,p_2,-p_1)$ may be cast in the form
\mbox{$\Gamma^a_{\mu}(q,p_2,-p_1) = gt^a \Gamma_{\mu}(q,p_2,-p_1)$}, where $g$ is the gauge coupling and
$t^a$ are the SU(3) generators in the fundamental representation. $\Gamma_{\mu}(q,p_2,-p_1)$
may be then separated into two distinct pieces, 
\be
\Gamma_{\mu}(q,p_2,-p_1)=\Gamma^{\STI}_{\mu}(q,p_2,-p_1)+\Gamma^{\Tr}_{\mu}(q,p_2,-p_1)\,,
\label{split}
\ee
where $\Gamma^{\Tr}_{\mu}$ is transverse
with respect to the momentum $q^{\mu}$ carried by the gluon,  
\be
q^{\mu}\Gamma^{\Tr}_{\mu}(q,p_2,-p_1)=0\,, 
\ee 
while the first piece saturates the fundamental STI given by 
\be
 q^{\mu}\Gamma^{\STI}_{\mu}(q,p_2,\!-p_1\!)\! = \!F(q)[S^{-1}\!(p_1)\! H\!(q,p_2,\!-p_1\!)\! -\!{\overline H}\!(\!-q,p_1,\!-p_2\!)S^{-1}\!(p_2)] \,.
\label{STI}
\ee
In the STI above,  
$F(q)$ is the dressing function of the full ghost propagator, $D(q)= iF(q)/q^2$, \mbox{$H$} is the quark-ghost scattering kernel,
shown diagrammatically in the second line of Fig.~\ref{fig:system}, while 
\mbox{${\overline {H}}$} is its ``conjugate'', whose relation to \mbox{$H$} is explained in detail 
in~\cite{Aguilar:2016lbe}. Note that the color structure has been factored out, setting $H^{a}=-gt^a H$. 
The  most general tensorial decompositions of $H$ and  $\overline H$  read~\cite{Davydychev:2000rt,Aguilar:2010cn,Aguilar:2016lbe}
\begin{align}
H(q,p_2,-p_1)=X_0\mathbb{I}+X_1\slashed{p}_1+X_2\slashed{p}_2+X_3\widetilde{\sigma}_{\mu\nu}p_1^{\mu}p_2^{\nu}\,,\nonumber \\
\overline{H}(-q,p_1,-p_2)=\overline{X}_0\mathbb{I}+\overline{X}_2\slashed{p}_1+\overline{X}_1\slashed{p}_2+\overline{X}_3\widetilde{\sigma}_{\mu\nu}p_1^{\mu}p_2^{\nu}\,, 
\label{Hdecomp}
\end{align}
where $\widetilde\sigma_{\mu\nu} \equiv \frac{1}{2}[\gamma_{\mu},\gamma_{\nu}]$, and  we have introduced the  compact notation for the form factors \mbox{$X_i:= X_i(q^2,p_2^2,p_1^2)$} and \mbox{$\overline{X}_i:= X_i(q^2,p_1^2,p_2^2)$}. At tree-level, $X^{(0)}_0=\overline{X}^{(0)}_0=1$ and \mbox{$X^{(0)}_i=\overline{X}^{(0)}_i=0$} for \mbox{$i \geq 1$}.

Next, we can write the most general Lorentz decomposition for $\Gamma^{\STI}_{\mu}$ as
\be
\Gamma^{\STI}_{\mu}(q,p_2,\!-p_1\!)= 
  L_1 \gamma_{\mu}
+ L_2 (\slashed{p}_1 - \slashed{p}_2)(p_1-p_2)_{\mu} 
+ L_3 (p_1-p_2)_{\mu} 
+ L_4 \tilde\sigma_{\mu\nu}(p_1-p_2)^{\nu} \,,
\label{Li}
\ee
where $L_i:=L_i(q^2, p_2^2,p_1^2)$ are the quark-gluon form factors. 

It is clear that with the help of the Eq.~(\ref{STI})
the  form factors $L_i$, appearing in Eq.~\eqref{Li}, may be expressed 
in terms of $A$, $B$, $F$, $X_i$, and $\overline{X}_i$. Factoring
out the common $F(q)$, it is convenient to define  
\mbox{$L_i:=F(q)\overline{L}_i/2$}, which leads us to  
\bea
\overline{L}_1 &=& 
A(p_1)[X_0 - (p_1^2+p_1\!\cdot\!p_2)X_3] 
+ A(p_2)[{\overline X}_0 -(p_2^2 + p_1\!\cdot\!p_2){\overline X}_3]
\nonumber\\
&+& B(p_1)(X_2-X_1) + B(p_2)({\overline X}_2-{\overline X}_1);
\nonumber\\
\overline{L}_2 &=& \frac{1}{(p_1^2 - p_2^2)} \left\{
A(p_1)[X_0 + (p_1^2 - p_1\!\cdot\!p_2)X_3] 
- A(p_2)[{\overline X}_0 +(p_2^2-p_1\!\cdot\!p_2){\overline X}_3]\right\}
\nonumber\\
&-&
\frac{1}{(p_1^2 - p_2^2)} \left\{ B(p_1)(X_1+X_2) - B(p_2)({\overline X}_1+{\overline X}_2)\right\};
\nonumber\\
\overline{L}_3 &=&  \frac{2}{p_1^2 - p_2^2}
\left\{  
A(p_1) \left( p_1^2 X_1 + p_1\!\cdot\!p_2 X_2 \right)
- A(p_2) \left( p_2^2 {\overline X}_1 +p_1\!\cdot\!p_2 {\overline X}_2\right)
- B(p_1)X_0 + B(p_2){\overline X}_0\right\};
\nonumber\\
\overline{L}_4 &=& 
A(p_1) X_2 - A(p_2) {\overline X}_2 - B(p_1) X_3 + B(p_2){\overline X}_3 \,.
\label{expLi}
\eea

Setting in Eq.~\eqref{expLi} 
tree level values for the $X_i$  and $\overline{X}_i$, we obtain the form factors of
the ``minimally non-abelianized'' Ball-Chiu  vertex, 
$\Gamma^{\FBC}_{\mu} = F(q) \Gamma^{\rm \s{BC}}_{\mu}$, given by~\cite{Fischer:2003rp, Aguilar:2010cn,Aguilar:2016lbe}, 
\begin{align}\label{FBC}
L_1^{\FBC} &= F(q)\frac{[A(p_1)+A(p_2)]}{2}\,,\quad\quad L_2^{\FBC} = F(q)\frac{[A(p_1)- A(p_2)]}{2(p_1^2 - p_2^2)},  \nonumber \\
L_3^{\FBC} &= - F(q)\frac{[B(p_1)- B(p_2)]}{p_1^2 - p_2^2} \,, \qquad  L_4^{\FBC} = 0 \,.
\end{align}
To proceed, we will  insert into Eq.~\eqref{gap} the dressed quark-gluon vertex of Eq.~\eqref{Li}, defining \mbox{$p_1 = p$}  and \mbox{$p_2 = k$}. It is important
to keep in mind that the expressions for the form
factors \mbox{$L_i=F(q)\overline{L}_i/2$} in terms of the  $X_i$ are given by Eq.~\eqref{expLi}. Then, taking appropriate traces and applying the usual
rules for going to Euclidean space~\cite{Aguilar:2010cn}, we derive the following expressions for the integral equations satisfied by $A(p)$ and $B(p)$, 
\bea\label{gAB}
p^2A(p)  &=& Z_{\rm{\s F}} p^2 + Z_1 4\pi C_{F}\alpha_s\!\int_{k} {\mathcal K}_{\rm{\s A}}(k,p)\Delta(q)F(q)\, , \nonumber\\
 B(p)  &=& Z_1 4\pi C_{F} \alpha_s\! \int_{k} {\mathcal K}_{\rm{\s B}}(k,p)\Delta(q)F(q)\,,
\eea 
where \mbox{$\alpha_s = g^2(\mu)/4\pi$}, and  we have introduced the kernels 
\bea
{\mathcal K}_{\rm{\s A}}(k,p)&=&\left\{\frac{3}{2} (k \!\cdot\!p)\overline{L}_{1} -[\overline{L}_{1} -(k^2+p^2) \overline{L}_{2}]h(p,k) \right\}{\mathcal Q}_{\rm{\s A}}(k) \nonumber\\ 
&-& \left\{\frac{3}{2}p \!\cdot\!(k+p)\overline{L}_{4} +(\overline{L}_{3} - \overline{L}_{4})h(p,k)\right\}{\mathcal Q}_{\rm{\s B}}(k)\,,\nonumber\\ 
{\mathcal K}_{\rm{\s B}}(k,p)&=&\left\{\frac{3}{2} k \!\cdot\!(k+p)\overline{L}_{4} - (\overline{L}_{3}+\overline{L}_{4})h(p,k) \right\}{\mathcal Q}_{\rm{\s A}}(k) \nonumber\\ 
 &+& \left\{\frac{3}{2}\overline{L}_{1}  - 2 h(p,k)\overline{L}_{2}\right\}{\mathcal Q}_{\rm{\s B}}(k)\,,
\label{kernelAB}
\eea
with the functions $h(p,k)$ and ${\mathcal Q}_{\rm{\s f}}(k)$ defined as 
\be
h(p,k) := \frac{k^2p^2-(k\!\cdot\!p)^2}{q^2} \,,
\label{hf}
\ee
and 
\bea
{\mathcal Q}_{\rm{\s f}}(k) := \frac{f(k)}{A^2(k)k^2+B^2(k)}\,,
\label{ratio}
\eea
where $f(k)$, appearing in the numerator of Eq.~\eqref{ratio}, can be either $A(k)$ or $B(k)$, depending on the index of  ${\mathcal Q}$.

Clearly,  the kernels  ${\mathcal  K}_{\rm{\s  A}}$  and
${\mathcal   K}_{\rm{\s   B}}$ that enter in 
Eqs.~\eqref{gAB} depend   on  the  various   $\overline{L}_i$,  which
ultimately will couple  the functions $A(p)$ and $B(p)$  with the four
integral equations for the form factors  $X_i$, to be presented in 
Eq.~\eqref{generalx}.  However, as we  explain in the next subsection,
before  proceeding  to  the  solution of  the  system,  an  additional
important  approximation  needs to  be  implemented  at the  level  of
Eqs.~\eqref{gAB}.

\subsection{\label{sec:renor} Approximate renormalization and the anomalous dimension of ${\mathcal{M}}(p)$}

It is relatively straightforward to establish that the 
  STI of Eq.~\eqref{STI}
 imposes the relation \mbox{$Z_1 = Z_{\rm{\s c}}^{-1}Z_{\rm{\s F}} Z_{\rm{\s H}}$}, where $Z_{\rm{\s c}}$ and $Z_{\rm{\s H}}$ are the renormalization constants 
 of the ghost propagator and the quark-ghost scattering kernel, respectively.  Now, we recall that, in the Landau gauge, both the quark self-energy and the quark-ghost  kernel are finite at one-loop~\cite{Nachtmann:1981zg}; thus, at that order, $Z_{\rm{\s F}}=Z_{\rm{\s H}}=1$, and, therefore, $Z_1=Z_{\rm{\s c}}^{-1}$.
Imposing the above relations on Eqs.~\eqref{gAB}, we obtain the approximate version 
\bea
p^2A(p)  &=& p^2 + Z_{\rm{\s c}}^{-1} 4\pi C_{F}\alpha_s\!\int_{k} {\mathcal K}_{\rm{\s A}}(k,p)\Delta(q)F(q)\, , \nonumber\\
 B(p)  &=& Z_{\rm{\s c}}^{-1} 4\pi C_{F} \alpha_s\! \int_{k} {\mathcal K}_{\rm{\s B}}(k,p)\Delta(q)F(q)\,.
\label{rsenergy1}
\eea%

Even with these approximations, the presence of $Z_{\rm{\s c}}^{-1}$ in front of the corresponding
integrals complicates the analysis,
especially in a non-perturbative setting~\cite{Brown:1989hy,Curtis:1990zs,Curtis:1993py,Bloch:2001wz,Bloch:2002eq}.
It is well-known that, in general, such multiplicative renormalization constants are instrumental for 
the systematic cancellation of overlapping divergences, whose complete implementation  
hinges, in addition, on the inclusion of crucial contributions stemming from the transverse
parts of the vertices involved (in our case, $\Gamma^{\Tr}_{\mu}$).
From the perturbative point of
view, several of the aforementioned issues have been studied in detail in the context of the electron
propagator in QED~\cite{Kizilersu:2009kg}, 
and even though the levels of technical complexity are high, they are considered
to be well-understood.
On the other hand, these 
cancellations are far more difficult to identify and enforce non-perturbatively, 
even if a reasonable approximation
of $\Gamma^{\Tr}_{\mu}$ is furnished. Given that in the present analysis the
term $\Gamma^{\Tr}_{\mu}$ is completely undetermined, and is set identically to zero,
the possibility of a {\it bona fide} cancellation of the overlapping divergences is
excluded from the outset.

A typical manifestation of the mismatches induced 
when implementing the usual simplification $Z_c^{-1}=1$
(or directly $Z_1=1$) is the failure of ${\mathcal{M}}(p)$ 
to display the correct anomalous dimension in the deep ultraviolet.
Specifically, the  asymptotic behavior of ${\mathcal{M}}(p)$ 
at one-loop is given  
by~\cite{Politzer:1976tv,Miransky:1981rt,Roberts:1994dr} 
\bea
{\mathcal M}_{\rm{\s{UV}}}(p) = \frac{C}{p^2}
\left[ \ln\left(\frac{p^2}{\Lambda^2}\right)\right]^{\gamma_f -1} \,,
\label{fituv}
\eea
where $C$ is a constant with mass dimension $[M]^3$, $\gamma_f =12/(11C_{\rm A}-2n_f)$ is 
the mass anomalous dimension, and $n_f$ is the number of active quark flavors. Instead, if the aforementioned approximation is implemented, the asymptotic behavior of the quark mass obtained from the resulting gap equation 
has the wrong value for $\gamma_f$,  given by  $\gamma_f= 48/(35C_{\rm A} -8n_f)$.

A simple remedy to this problem has been put forth in~\cite{Aguilar:2010cn}, which is similar in spirit to an
earlier proposal presented in~\cite{Fischer:2003rp}. 
Specifically, one carries out 
the substitution 
\be
Z_c^{-1}{\mathcal K}_{\rm{\s {A,B}}}(p,k) \to
 {\mathcal K}_{\rm{\s{A,B}}}(p,k){\mathcal C}(q),
\label{replace}
 \ee
 where  the function ${\mathcal C}(q)$ should  display the appropriate ultraviolet characteristics to convert the  product 
\bea\label{RGI}
{\mathcal R}(q)= \alpha_s(\mu)\Delta(q,\mu)F(q,\mu){\mathcal C}(q,\mu)\,,
\eea
into a renormalization-group invariant (RGI) ($\mu$-independent) combination, at least at one-loop.

Focusing on the function ${\mathcal C}(q)$, the requirement that ${\mathcal R}(q)$ be RGI 
fixes its ultraviolet behavior; specifically, for large $q^2$, the inverse of ${\mathcal C}(q)$
must behave as 
\bea
{\mathcal C}_{\rm{\s{UV}}}^{-1}(q) = 1+\frac{9C_{\rm{\s A}} \alpha_s}{48\pi}\ln\left(\frac{q^2}{\mu^2}\right)\,,
\label{FUV}
\eea  
where $C_{\rm{\s A}}$ is the eigenvalue of the Casimir operator in the adjoint representation.
However, the low-energy completion of ${\mathcal C}(q)$ remains undetermined, leading to the
necessity of introducing specific {\it Ans\"atze} for it. 

The ghost dressing function $F(q)$ is the simplest quantity that
fulfills \eqref{FUV} and, due to high-quality lattice simulations and extensive
studies in the continuum, is quite accurately known in the entire range of Euclidean momenta.
However, in the present work we will mainly focus on   
an alternative quantity that conforms with the aforementioned requirements,
and, in addition, displays a relative enhancement with respect to $F(q)$ in the region of momenta
that is particularly relevant for chiral symmetry breaking. Specifically, we will employ 
the so-called ``ghost-gluon'' mixing self-energy, denoted by $1+ G(q)$,
which is a crucial ingredient in contemporary application of the pinch technique~\cite{Binosi:2009qm,Aguilar:2008xm,Binosi:2014aea,Binosi:2016nme},
and coincides (in the Landau gauge) with the well-known
Kugo-Ojima function~\cite{Kugo:1979gm,Kugo:1995km,Grassi:2004yq,Fischer:2006ub,Aguilar:2009pp}. 
The quantity $[1+G(q)]^{-1}$ has precisely the same ultraviolet behavior stated in \eqref{FUV},
and SDE and lattice studies furnish its form for low and intermediate momenta (see Fig.~\ref{func_ren}); 
in fact, by virtue of an exact identity valid in the Landau gauge,
$[1+G(0)]^{-1} = F(0)$~\cite{Aguilar:2009nf}.

An accurate fit of
$1+G(q)$, valid for the entire range of Euclidean momenta, is given by 
\be
 1+G(q) = 1+\frac{9C_{\rm{\s A}}\alpha_s}{48\pi}I(q)\ln\left(\frac{q^2+\rho_3 m^2(q)}{\mu^2}\right) \,, 
 \label{1G}
 \ee 
with 
 \bea
m^2(q)&=&\frac{m^4}{q^2+ \rho_2 m^2}\,, \nonumber \\
I(q)&=& 1+ D \exp{\left(-\frac{\rho_4 q^2}{\mu^2}\right)}\,,
\eea
where
\mbox{$m^2=0.55\,\mbox{GeV}^2$},  
\mbox{$\rho_2=0.60$}, 
\mbox{$\rho_3=0.50$},
\mbox{$\rho_4=2.08$},
\mbox{$\alpha_s=0.22$},
\mbox{$D=3.5$}, and  
\mbox{$\mu=4.3$\,GeV}.

In the general analysis presented in the following section,
we will consider three particular models 
for ${\mathcal C}(q)$; the first two  have the function  $[1+G(q)]^{-1}$ as their principal ingredient, while the third is simply $F(q)$ itself. 
Of course, these Ans\"atze are to be understood 
as representative cases of a wider range of qualitatively similar, but technically
more involved, realizations~\footnote{For example, if ${\mathcal C}(q)$ originates ultimately 
from $\Gamma^{\Tr}_{\mu}$,
it would be reasonable to expect its dependence on $k$, $p$, and $\theta$ 
to be more complicated than simply $q^2= (k-p)^2$.}.
  
Specifically,  
\bea
 {\mathcal C}_1(q)&=& [1+G(q)]^{-1}\,, \nonumber \\ 
{\mathcal C}_2(q)&=&\frac{q^2}{q^2 + a_{1}}\left[1+ \exp\left(-\frac{a_2 q^2}{\mu^2}\right)\right][1+G(q)]^{-1}\,, \nonumber \\ 
{\mathcal C}_3(q)&=& F(q) \,,
\label{rfunc}
\eea
where \mbox{$a_1=0.13\,\mbox{GeV}^2$}  and $a_2=50$. Note that $F^{-1}(q)$
can be also expressed by the same functional form given in Eq.~\eqref{1G}, where 
the corresponding fitting parameters are \mbox{$m^2=0.55\,\mbox{GeV}^2$},  
\mbox{$\rho_2=2.57$}, \mbox{$\rho_3=0.50$}, \mbox{$\rho_4=3.83$}, and \mbox{$D=2.24$}.

By construction,  the three {\it Ans\"atze} display the same asymptotic behavior, 
and their perturbative tails merge into each other approximately in the region of \mbox{$3$ GeV} [see Fig.~\ref{func_ren}]. In addition, one can see that ${\mathcal C}_3(q)$ is more suppressed than ${\mathcal C}_1(q)$ and ${\mathcal C}_2(q)$ in the range of [400~MeV, 2~GeV]. On the other hand, the main difference between the first two {\it Ans\"atze}  appears
below approximately \mbox{700~MeV};
thus, while ${\mathcal C}_1(q)$ grows monotonically and finally
saturates at the value $F(0)$, ${\mathcal C}_2(q)$ drops
rapidly and vanishes at the origin.

Finally, carrying out the replacement given in Eq.~\eqref{replace} into
Eq.~\eqref{gAB}, we obtain the
form of the gap equation that will be used in what follows; in particular,  
\bea\label{gAB1}
p^2A(p)  &=& p^2 + 4\pi C_{\rm{\s F}} \int_{k} {\mathcal K}_{\rm{\s A}}(k,p){\mathcal R}_i(q)\, , \nonumber\\
 B(p)  &=& 4\pi C_{\rm{\s F}}\int_{k} {\mathcal K}_{\rm{\s B}}(k,p){\mathcal R}_i(q)\,,
\eea 
where ${\mathcal R}_i(q)$ refers to the RGI product of Eq.~\eqref{RGI}, realized with ${\mathcal C}_i(q)$, for \mbox{$i=1,\,2 \;\mbox{or}\;3$}.

\subsection{\label{sec:syst} The equations for the $X_i$}

The starting point in deriving the dynamical equations governing the behavior
of the form factors $X_i$ is the diagrammatic representation 
of $H^{[1]}(q,k,-p)$ at the one-loop dressed approximation, shown 
in the second line of Fig.~\ref{fig:system}.

As we can see, the complete treatment of  $H^{[1]}(q,k,-p)$ requires  the previous 
knowledge of the full gluon-ghost vertex, $G_{\nu}$, and 
the complete quark-gluon vertex, $\Gamma_{\mu}$, including its transverse part.
In order to reduce the level of technical complexity,
we  will adopt the following approximations: 
\n{i} for the full gluon-ghost vertex we
simply use its tree-level value \mbox{$G_{\nu}^{(0)}=(p-l)_{\mu}$},
and \n{ii} $\Gamma_{\mu}$ is approximated 
by the component of $\Gamma^{\rm \s{BC}}_{\nu}$,
proportional to $\gamma_{\nu}$, namely the 
$L_1^{\rm \s{FBC}}$ of Eq.~\eqref{FBC} with $F(q)=1$.  

With the above simplifications, one has
\be
H^{[1]} = 1- \frac{1}{2}iC_{\rm{\s A}}g^2 \int_{l} \Delta^{\mu\nu}(l-k)G_{\mu}^{(0)}(p-l)
D(l-p)S(l)L_1^{\rm \s{BC}}(l-k,k,-l)\gamma_{\nu}\,.
\label{eqH}
\ee
Then, contracting the above equation with the projectors defined 
in Eq.~(3.9) of~\cite{Aguilar:2016lbe}\footnote{Note that in the convention of momenta
used in~\cite{Aguilar:2016lbe} we have $p_1 \to p$ and $p_2 \to k$.}, one obtains 
the following  set of expressions for the individual form factors \mbox{$X_i(q^2,k^2,p^2)$},
\begin{align}
\label{generalx}
X_0 &=1+ i \pi C_{\rm{\s A}}\alpha_s\!\int_l\mathcal{K}(p,k,l)A(l){\mathcal G}(k,q,l)\,,\nonumber\\
X_1 &=i\pi  C_{\rm{\s A}}\alpha_s\int_l\frac{\mathcal{K}(p,k,l)B(l)}{q^2h(p,k)}\left[k^2{\mathcal G}(p,q,l)-(p\cdot k){\mathcal G}(k,q,l)\right]\,,\nonumber\\
X_2 &=i \pi C_{\rm{\s A}}\alpha_s\int_l\frac{\mathcal{K}(p,k,l)B(l)}{q^2h(p,k)}\left[p^2{\mathcal G}(k,q,l)-(p\cdot k){\mathcal G}(p,q,l)\right] \,,\nonumber\\
X_3 &=-i \pi C_{\rm{\s A}}\alpha_s\int_l\frac{\mathcal{K}(p,k,l)A(l)}{q^2h(p,k)}\left[ k^2{\mathcal G}(p,q,l) - (p\cdot k){\mathcal G}(k,q,l)  - {\mathcal T}(p,k,l)  \right]\,,
\end{align}
where we have introduced the kernel
\begin{align}
\mathcal{K}(p,k,l)&=\frac{F(l-p)\Delta(l-k)[A(l)+A(k)]}{(l-p)^2[A^2(l)l^2-B^2(l)]}\,,
\label{kernelsH}
\end{align}
and the functions 
\bea
{\mathcal G}(r,q,l) &=& (r\cdot q)-\frac{[r\cdot(l -k)][q\cdot(l-k)]}{(l-k)^2} \,, \nonumber \\
{\mathcal T}(p,k,l) &=& (k\cdot q)[(p\cdot l) - (p\cdot k)] - (p\cdot q)[(k\cdot l)-k^2] \,.
\label{fgg}
\eea
The above expressions for $X_i$ 
are expressed in Minkowski space, and depend on  the three momenta $q$, $k$, and $p$.
The Euclidean version of~\eqref{generalx}
is given in Eq.~(3.21) of~\cite{Aguilar:2016lbe},
and is a function of $p^2$, $k^2$, and the angle $\theta$ between $p$ and $k$,  
{\it i.e.}, \mbox{$X_i(p^2, k^2,\theta)$}.

\section{\label{Num} Numerical Analysis}

In this section we present the numerical analysis and main results of the 
six coupled integral equations formed by $A(p)$, $B(p)$, and the four $X_i$, defined by Eqs.~\eqref{gAB1} and~\eqref{generalx}, respectively.

\subsection{\label{Num_inputs}Inputs}

\begin{figure}[!t]
\begin{minipage}[b]{0.45\linewidth}
\centering
\includegraphics[scale=0.35]{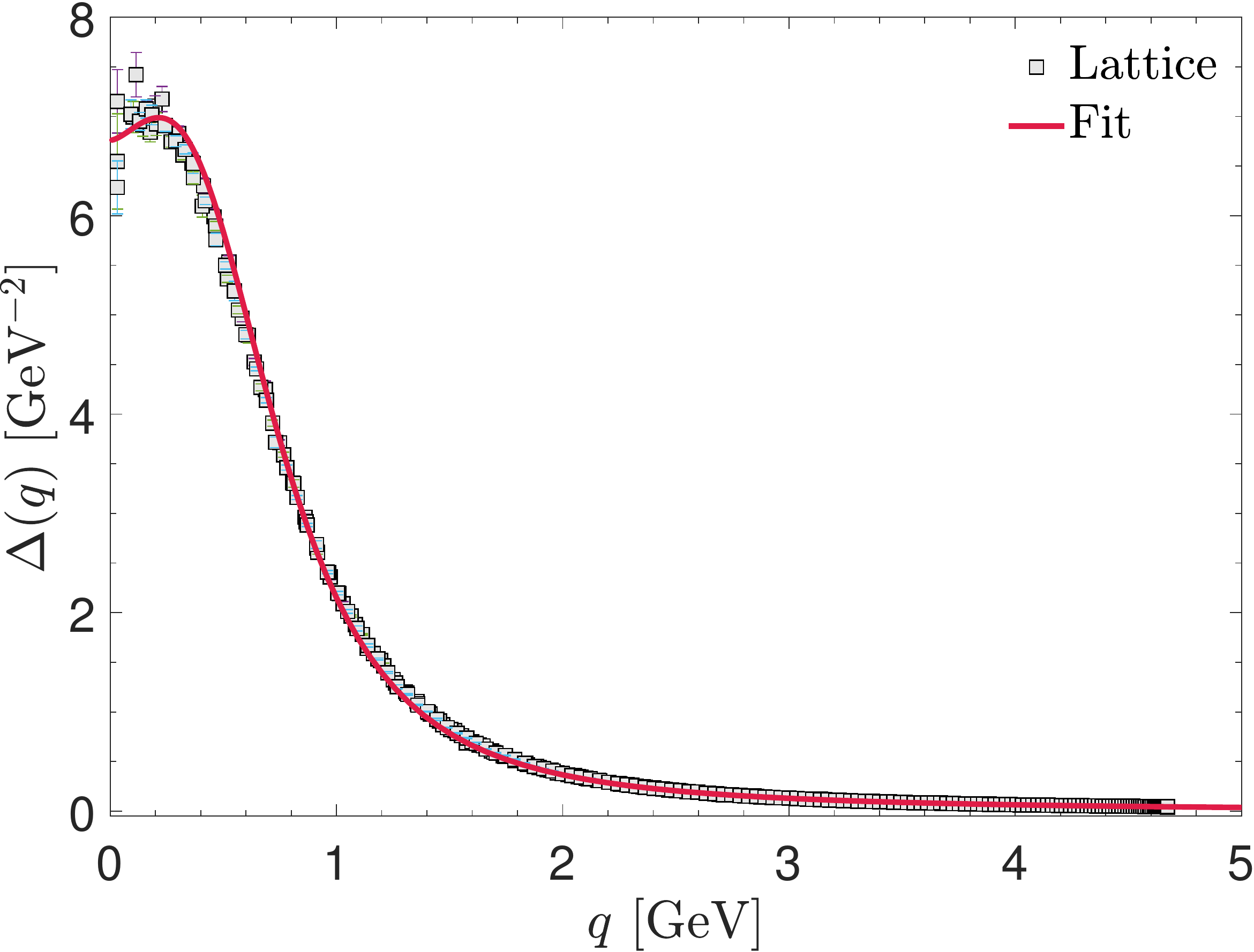}
\end{minipage}
\hspace{1.0cm}
\begin{minipage}[b]{0.45\linewidth}
\includegraphics[scale=0.35]{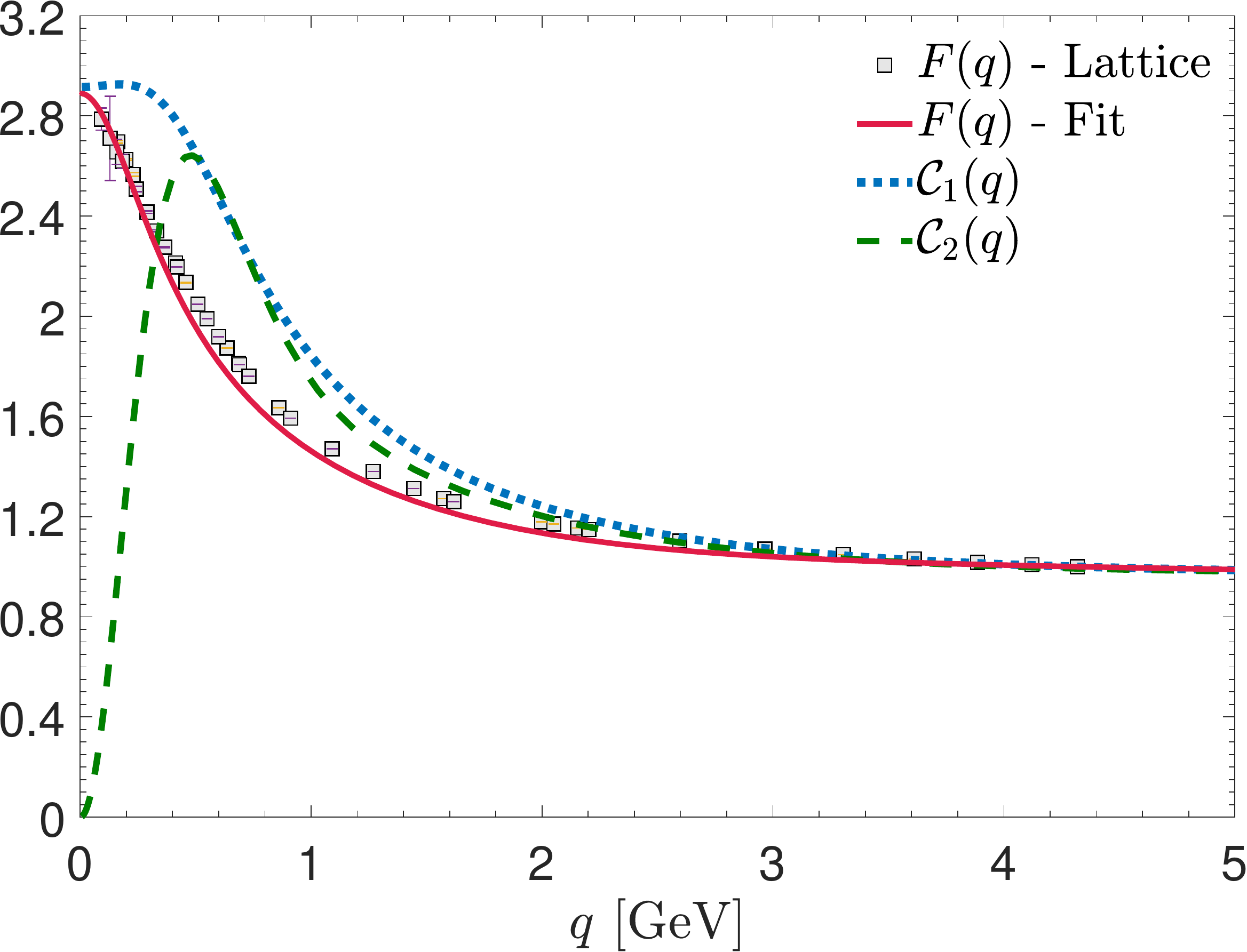}
\end{minipage}
\caption{\label{func_ren}  The  gluon propagator, $\Delta(q)$, (left panel, red continuous) and the ghost dressing function, $F(q)$, (right panel, red continuous), and the corresponding lattice data of~\cite{Bogolubsky:2007ud}.
In the right panel we also show the functions  
 ${\mathcal C}_1(q)$ (blue dotted), ${\mathcal C}_2(q)$ (green dashed), and
 ${\mathcal C}_3(q)=F(q)$ (red continuous) given by Eqs.~\eqref{rfunc}. 
 All functions are renormalized at \mbox{$\mu = 4.3$\,GeV}. }
\end{figure}

As can be observed from Eqs.~\eqref{gAB1} and~\eqref{generalx}, the numerical evaluation of  $A(p)$, $B(p)$, and $X_i$ requires
the knowledge of 
three additional quantities: {\it (i)} the gluon propagator, $\Delta(q)$, {\it (ii)} the ghost dressing function, $F(q)$, and  {\it (iii)} the function, ${\mathcal C}_i(q)$, appearing in the renormalized version of the gap equation~\eqref{gAB1}.    
Ideally one could consider an even more extended system of equations, where the six equations would be coupled to the
two additional SDEs that determine the momentum evolution of  $\Delta(q)$ and $F(q)$; however, the resulting complexity of such
an approach is very high. Instead, as was done in a series of earlier works~\cite{Aguilar:2010cn,Aguilar:2011ux,Aguilar:2011yb,Aguilar:2016lbe},  
we will employ for  $\Delta(q)$  and $F(q)$ appropriate fits reconstructed from  the lattice data of~\cite{Bogolubsky:2007ud}.
In the left panel of Fig.~\ref{func_ren} we show the lattice
data for $\Delta(q)$ and its corresponding fits (red continuous), renormalized at $\mu=4.3$ GeV.
We emphasize that these particular lattice results are ``quenched'', \ie do not incorporate the effects of dynamical
quark loops.  In addition, on the right panel of the same
figure,  we show ${\mathcal C}_1(q)$ (blue dotted), ${\mathcal C}_2(q)$ (green  dashed-dotted), and ${\mathcal C}_3(q)=F(q)$ (red continuous), all
given by  Eq.~\eqref{rfunc}.  Although  in the deep infrared and in the intermediate region of momenta the three curves display different behaviors, one  can clearly see that for values of \mbox{$q \gtrsim 3$ GeV}
they merge into each other, as discussed in the subsection \ref{sec:renor}.

The use of quenched lattice results merits
some additional clarifications, especially in view of the fact that
unquenched lattice data are also available in the literature; note, 
for instance, that the simulations of~\cite{Ayala:2012pb} yielded results for
both the gluon propagator and the ghost dressing function 
for  $N_f=2$ (two degenerate light quarks),
and  $N_f=2+1+1$ (two degenerate light quarks and two heavy ones). 

The main reason we refrain from using them is related with the fact that 
in such simulations
chiral symmetry is{ \it explicitly} broken due to the presence of a
non-vanishing current quark mass, $m_0(\mu)\neq 0$, whose inclusion
in the corresponding gap equation brings about nontrivial modifications. 
Specifically, the presence of a non-vanishing $m_0(\mu)$
introduces an  additional term $Z_m m_0(\mu)$
on the rhs of the equation for $B(p)$ given in~\eqref{gAB},
where $Z_m$ is the mass renormalization constant associated with $m_0(\mu)$.
The presence of this term complicates further 
the renormalization procedure of the gap equation.
To see that, we recall that the renormalization conditions in the momentum subtraction scheme (MOM) require that the renormalized $A(p)$ and $B(p)$  recover their tree level values at $\mu$, {\it i.e.}  $A(\mu)=1$, and 
$B(\mu)=m_0(\mu)$.  Then,  if one were to impose $Z_{\rm{\s F}}=1$ throughout, as was done in Sec.~\ref{sec:renor},
the renormalized $A(p)$ will not recover its tree level value at the renormalization point,
unless the contribution of the integral containing the kernel 
${\mathcal K}_{\rm{\s A}}(k,p)$ were vanishing.
Even though  
a ``hybrid'' treatment of $Z_{\rm{\s F}}$ could be adopted\footnote{For example, for the integral terms one may 
substitute \mbox{$Z_1{\mathcal K}_{\rm{\s {A,B}}}(p,k) \to
 {\mathcal K}_{\rm{\s{A,B}}}(p,k){\mathcal C}_i(q)$} as before, but treat ``subtractively'' 
the $Z_{\rm{\s F}}$ and $Z_m$ appearing in the ``tree-level'' terms.}, in order to
avoid these additional complications we use the quenched lattice results throughout.
 
 Let us finally mention that, notwithstanding the aforementioned difficulties,
 a rough estimate of the impact of the  “unquenching” effects in the form factors
of the quark-gluon vertex in some special kinematic limits was presented 
in~\cite{Aguilar:2014lha}; according to that analysis, 
the effects due to unquenching are relatively small, of the order of $10\%$.

\subsection{\label{Num_system}Numerical results for the coupled system}

With all external inputs defined, we proceed to solve the coupled system; 
note in particular that the form factors $X_i$ will be determined 
for general Euclidean kinematics.
Then, 
the vertex form factors $L_i$ will be obtained through direct substitution of the solutions into the Euclidean version of Eqs.~\eqref{expLi}.

 The coupled system of SDEs~\eqref{gAB1} and~\eqref{generalx} is solved iteratively. The  logarithmic grid is composed by $136$ different values of momenta $p^2$ in the range \mbox{$[ 5\times 10^{-5}\,\mbox{{GeV}}^2, 5\times 10^3\,\mbox{{GeV}}^2]$},  whereas the angular interval is subdivided uniformly into $25$ values from $0$ to $\pi$. The most costly task, the numerical evaluation of the multidimensional integrals, was tackled with an adaptative algorithm employing an 11th degree polynomial rule for the 3D integrals and a 13th degree rule for the 2D ones~\cite{Berntsen:1991:ADA:210232.210234}.
 
\begin{figure}[!h]
\begin{minipage}[b]{0.3\linewidth}
\centering
\includegraphics[scale=0.22]{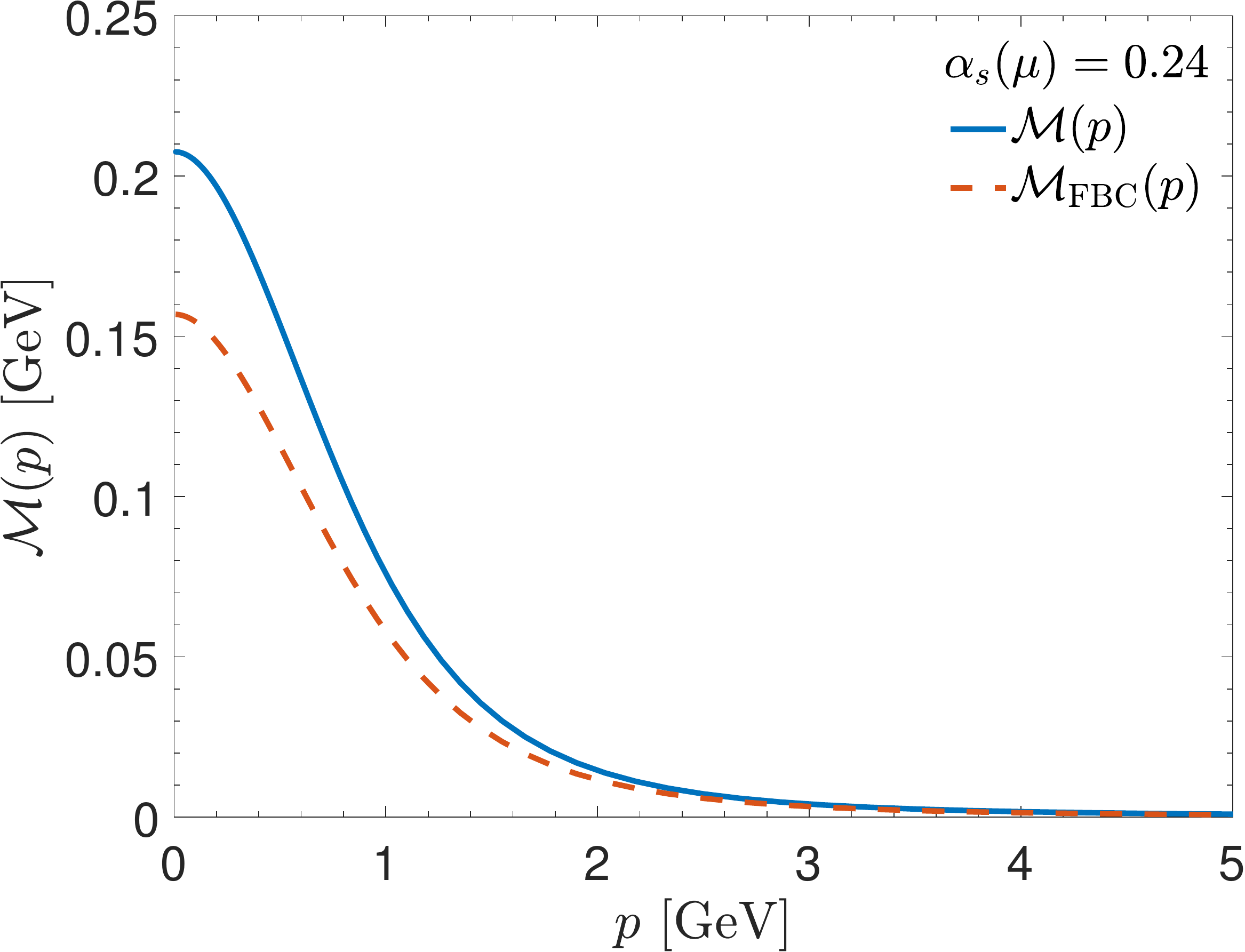}
\end{minipage}
\hspace{0.3cm}
\begin{minipage}[b]{0.3\linewidth}
\includegraphics[scale=0.22]{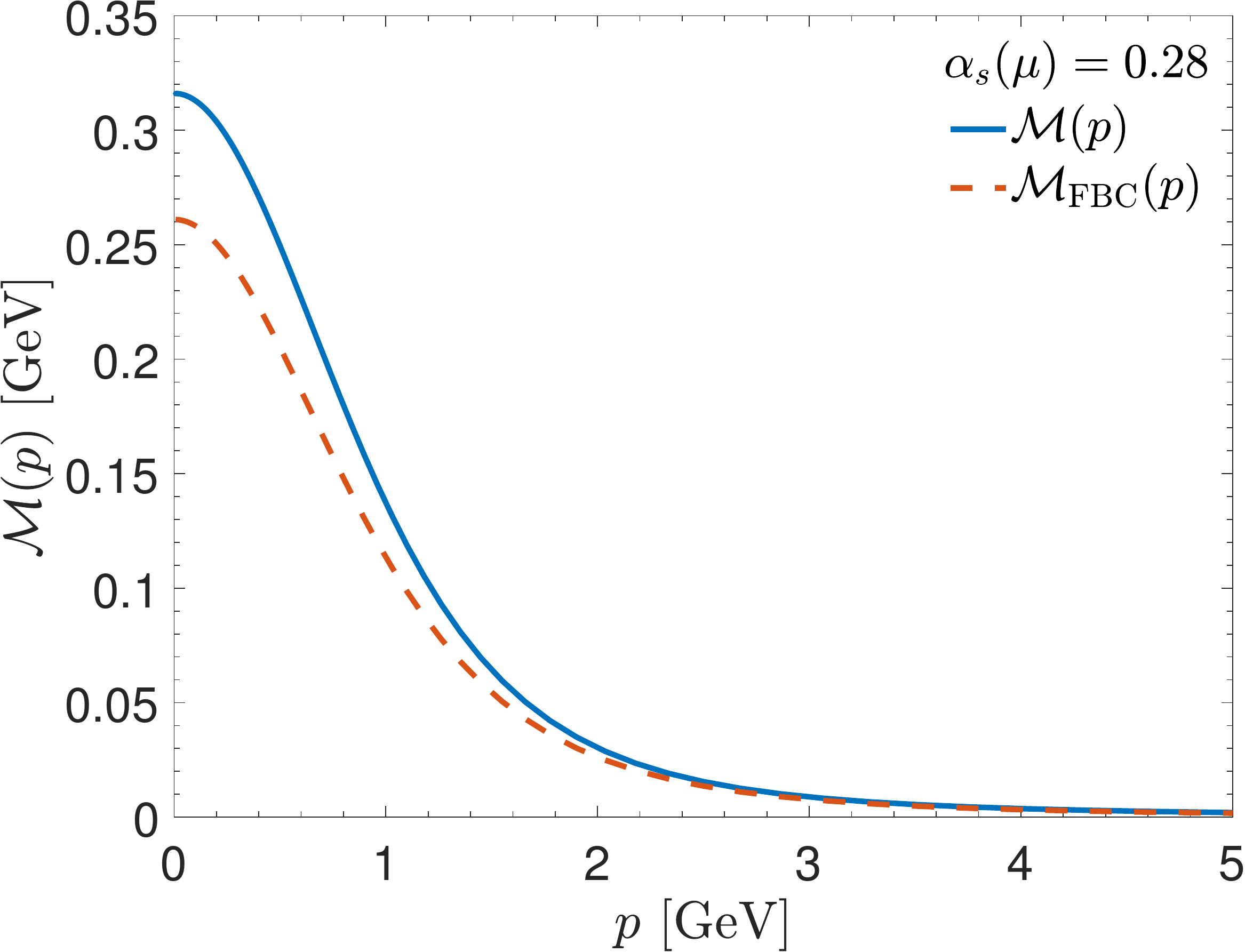}
\end{minipage}
\hspace{0.3cm}
\begin{minipage}[b]{0.3\linewidth}
\includegraphics[scale=0.22]{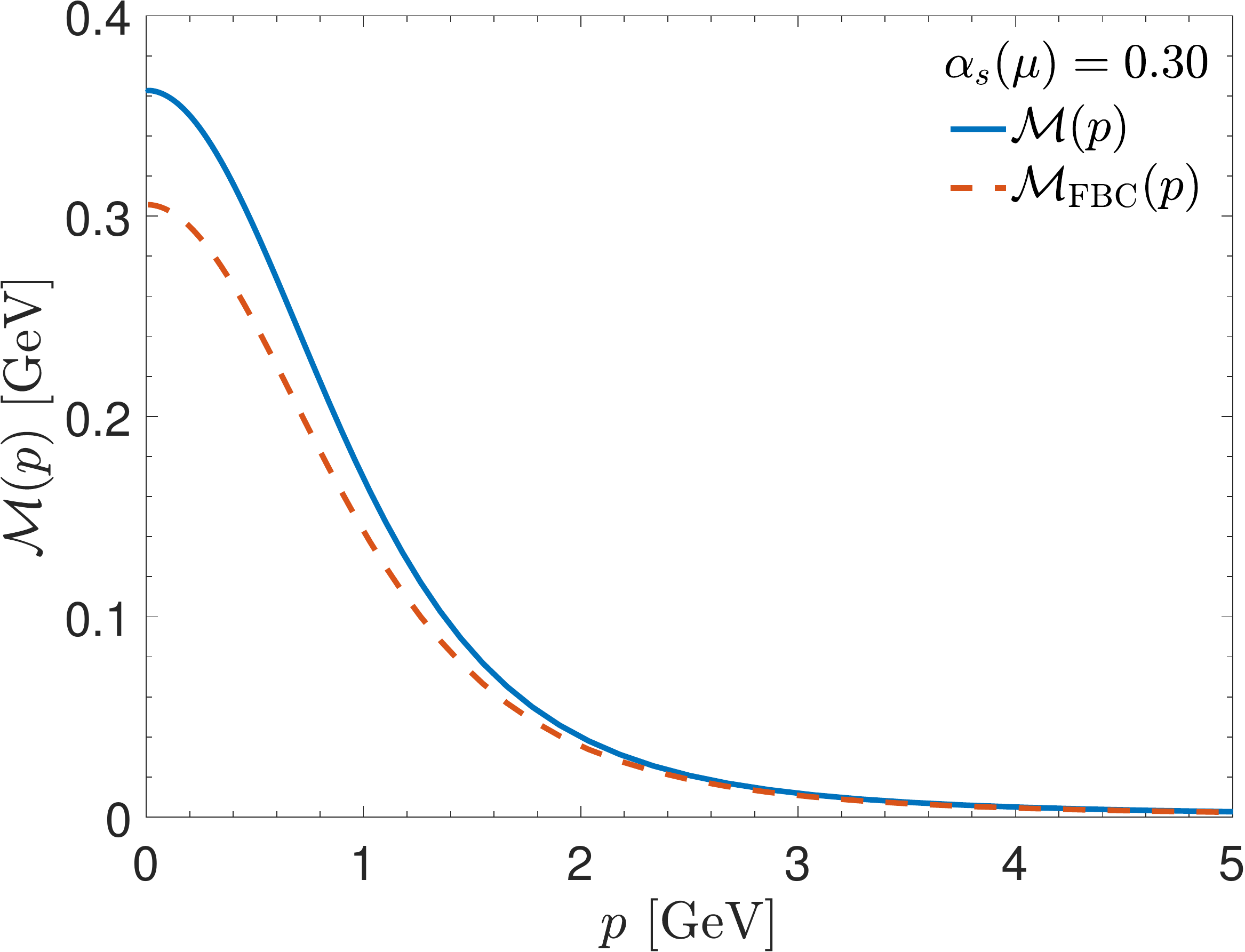}
\end{minipage}
\newline
\begin{minipage}[b]{0.3\linewidth}
\centering
\includegraphics[scale=0.22]{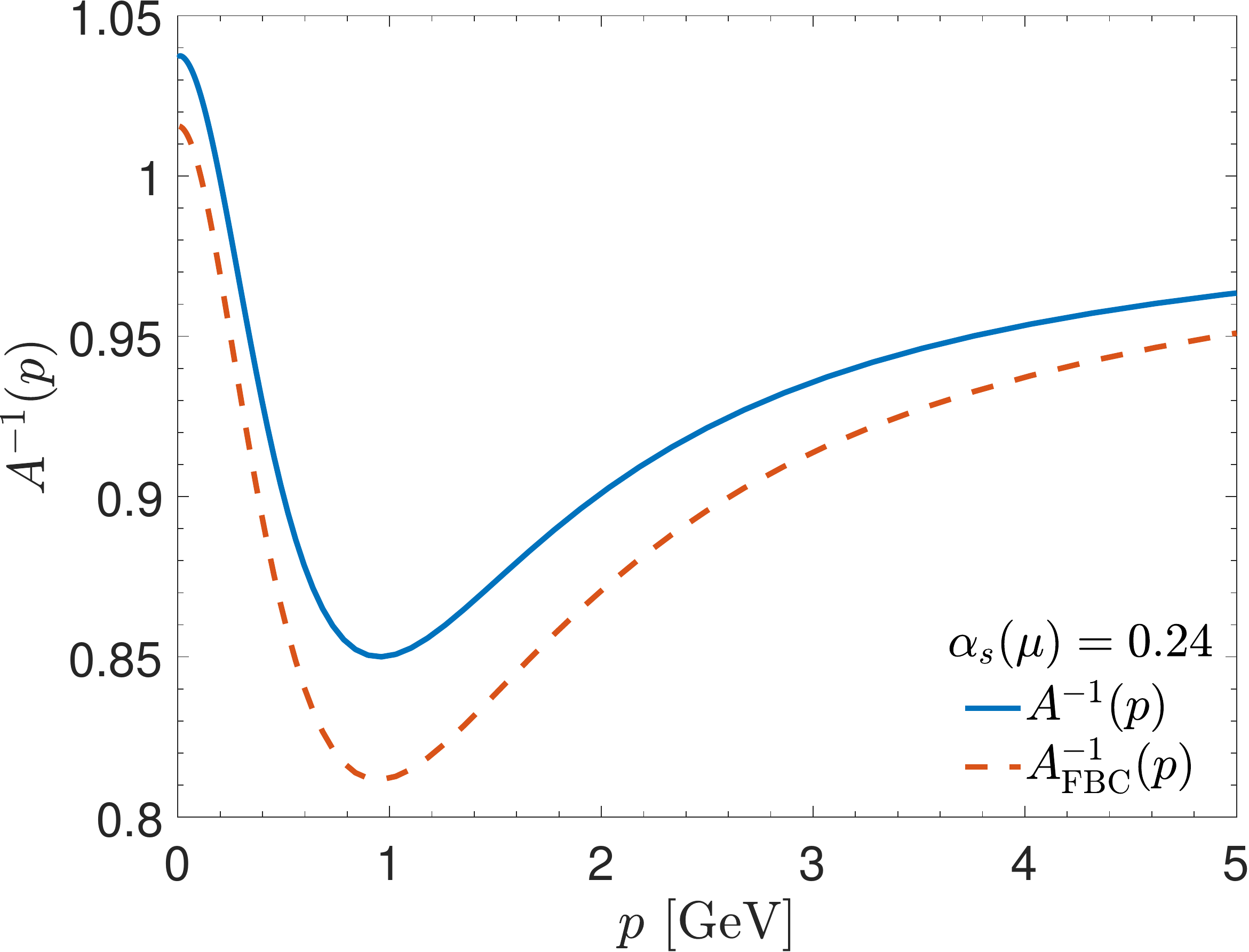}
\end{minipage}
\hspace{0.3cm}
\begin{minipage}[b]{0.3\linewidth}
\includegraphics[scale=0.22]{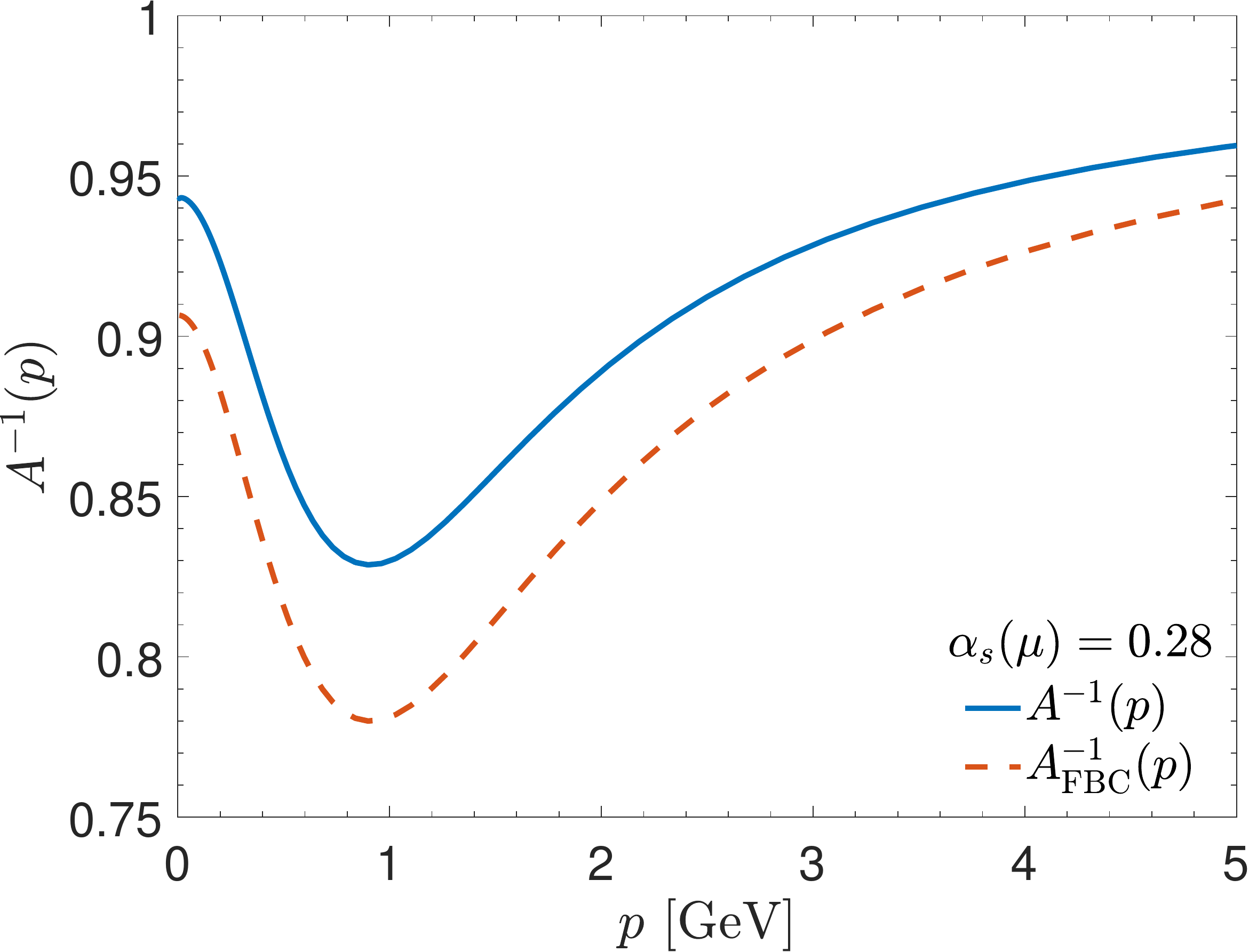}
\end{minipage}
\hspace{0.3cm}
\begin{minipage}[b]{0.3\linewidth}
\includegraphics[scale=0.22]{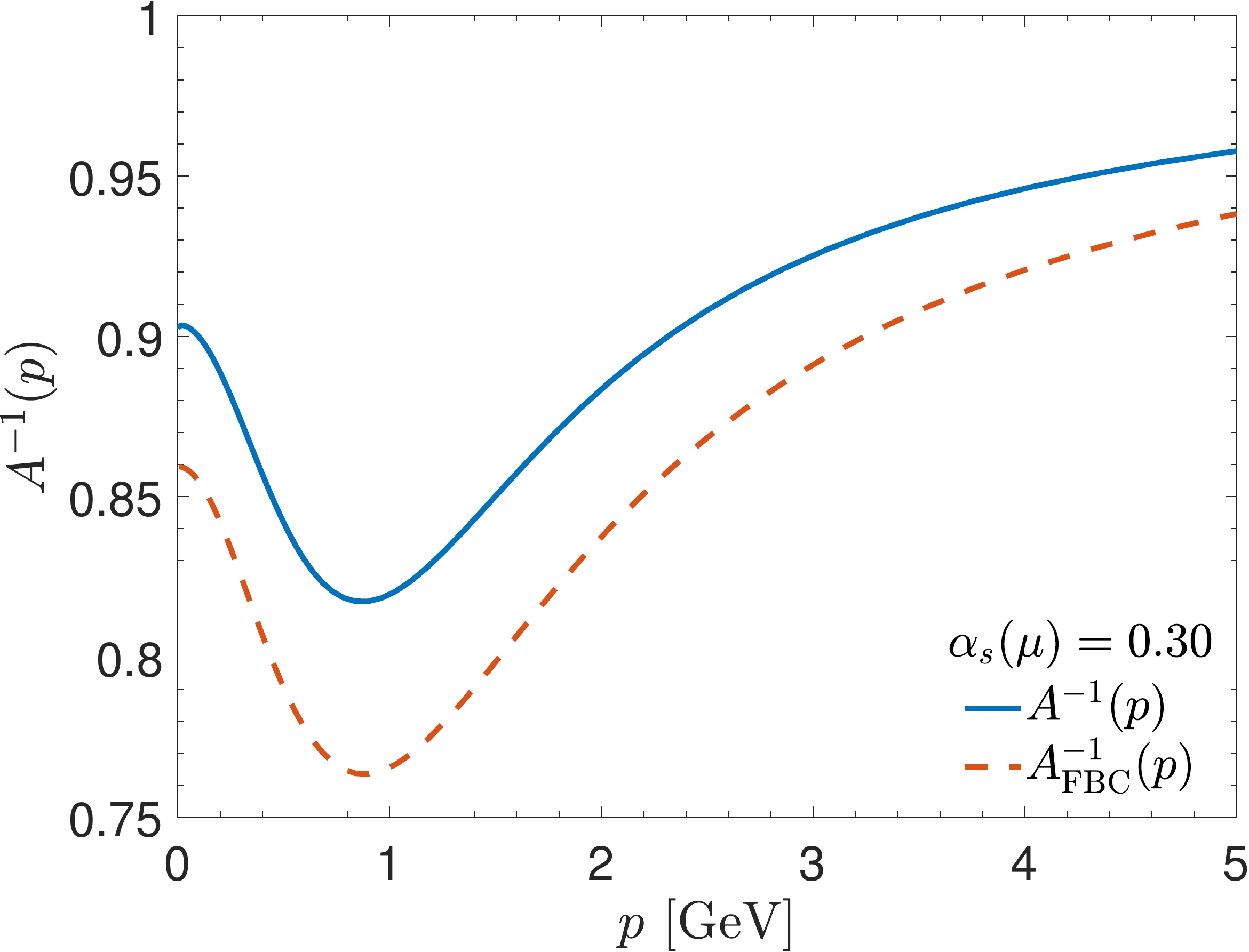}
\end{minipage}
\vspace{-0.3cm}
\caption{\label{mass_1G} Comparison of the ${\mathcal M}(p)$ (top panels) and the $A^{-1}(p)$ (bottom panels)
  obtained with $\Gamma^{\STI}_{\mu}$ (blue
  continuous curve) and those obtained using $\Gamma^{\FBC}_{\mu}$ (orange dashed curve).
  All curves were obtained with the ${\mathcal C}_1(q)$ of Eq.~\eqref{rfunc},
  and we used $\alpha_s=0.24$ (left panels), $\alpha_s=0.28$ (central panels), and $\alpha_s=0.30$ (right panels).}
\end{figure}
 
In Fig.~\ref{mass_1G} we show the numerical
results for two out of the six quantities determined in our coupled system. In particular, we show the  dynamical quark mass, ${\mathcal M}(p)$ (top panels), and the quark wave function, $A^{-1}(p)$ (bottom panels), obtained as solutions when we use  ${\mathcal C}_1(q)$ in  the RGI product ${\mathcal R}_1(q)$, defined
in the Eq.~\eqref{RGI}. The solutions were obtained for $\alpha_s=0.24$ (left panels), $\alpha_s=0.28$ (center panels), and $\alpha_s=0.30$ (right panels). 

\begin{figure}[!h]
\begin{minipage}[b]{0.40\linewidth}
\centering
\includegraphics[scale=0.37]{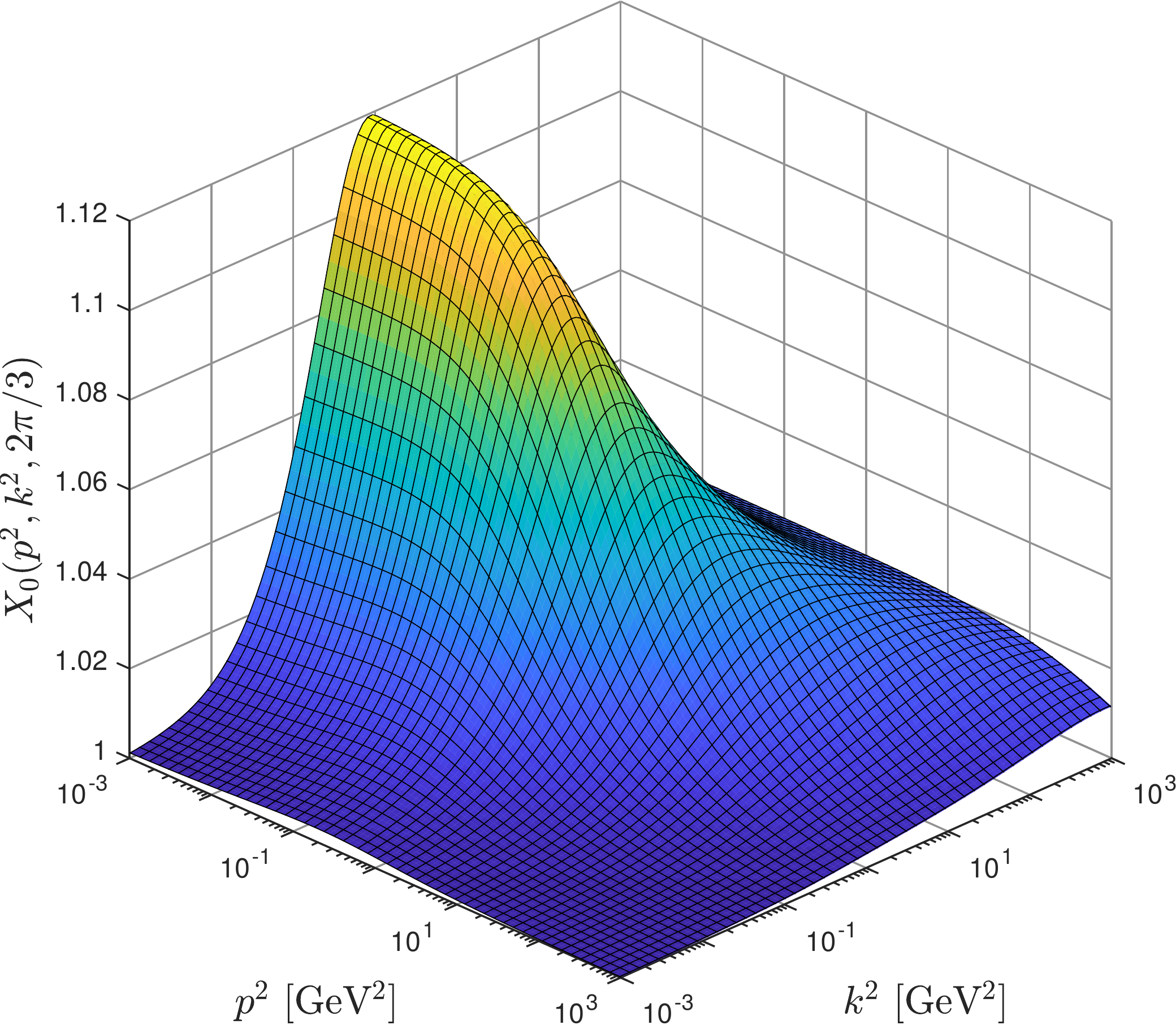}
\end{minipage}
\hspace{1.0cm}
\begin{minipage}[b]{0.45\linewidth}
\includegraphics[scale=0.37]{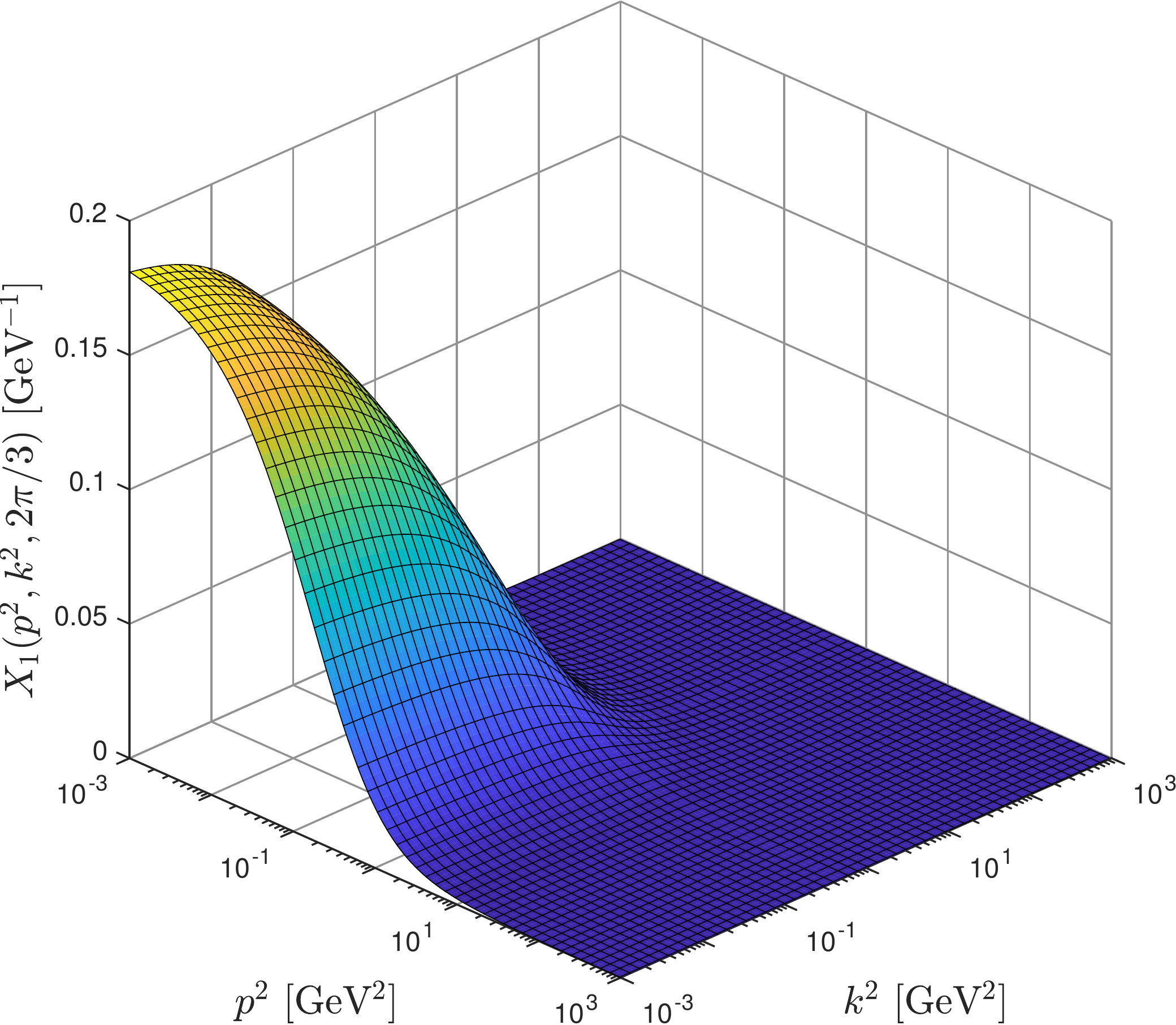}
\end{minipage}
\begin{minipage}[b]{0.40\linewidth}
\centering
\includegraphics[scale=0.37]{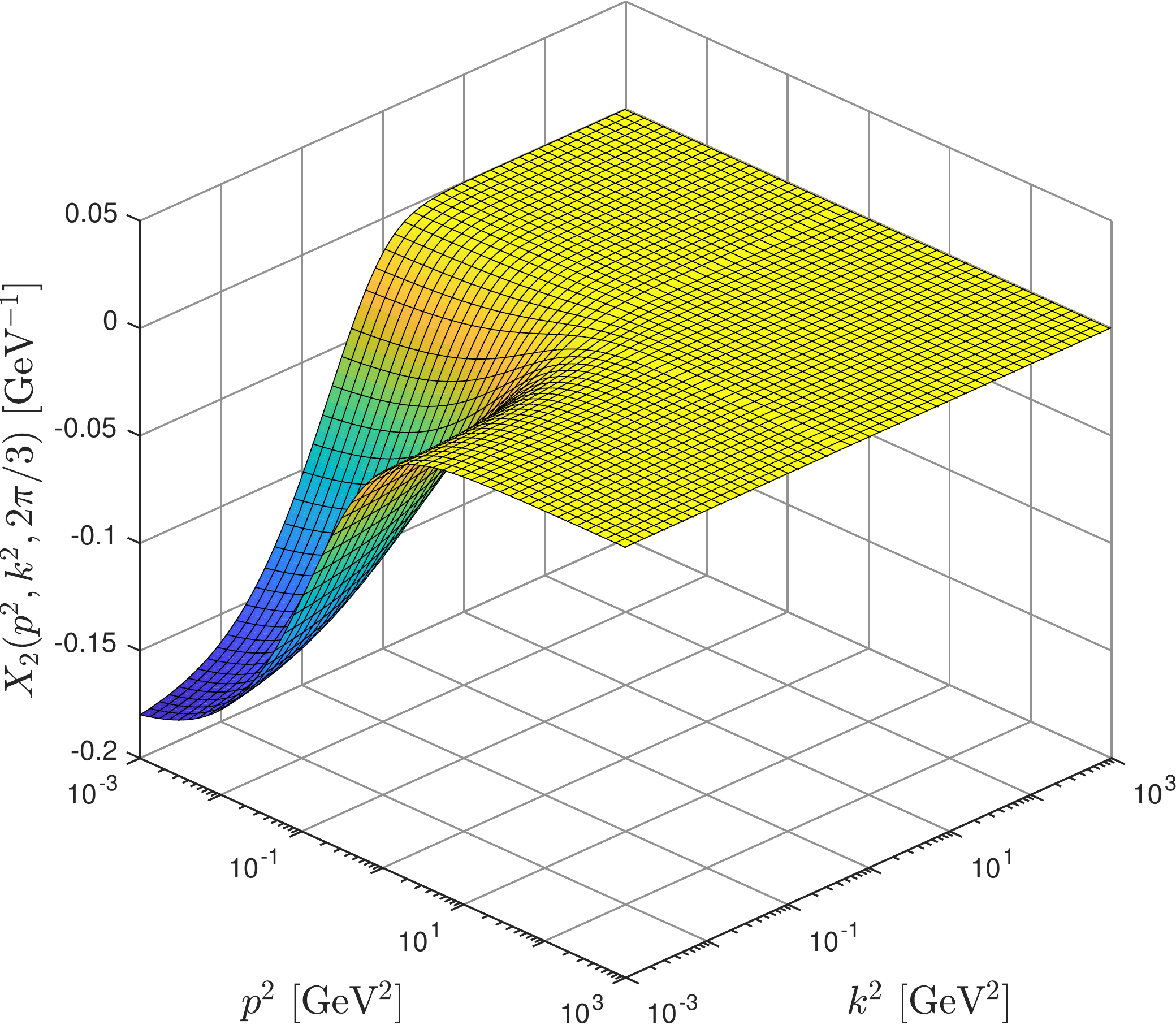}
\end{minipage}
\hspace{1.0cm}
\begin{minipage}[b]{0.45\linewidth}
\includegraphics[scale=0.37]{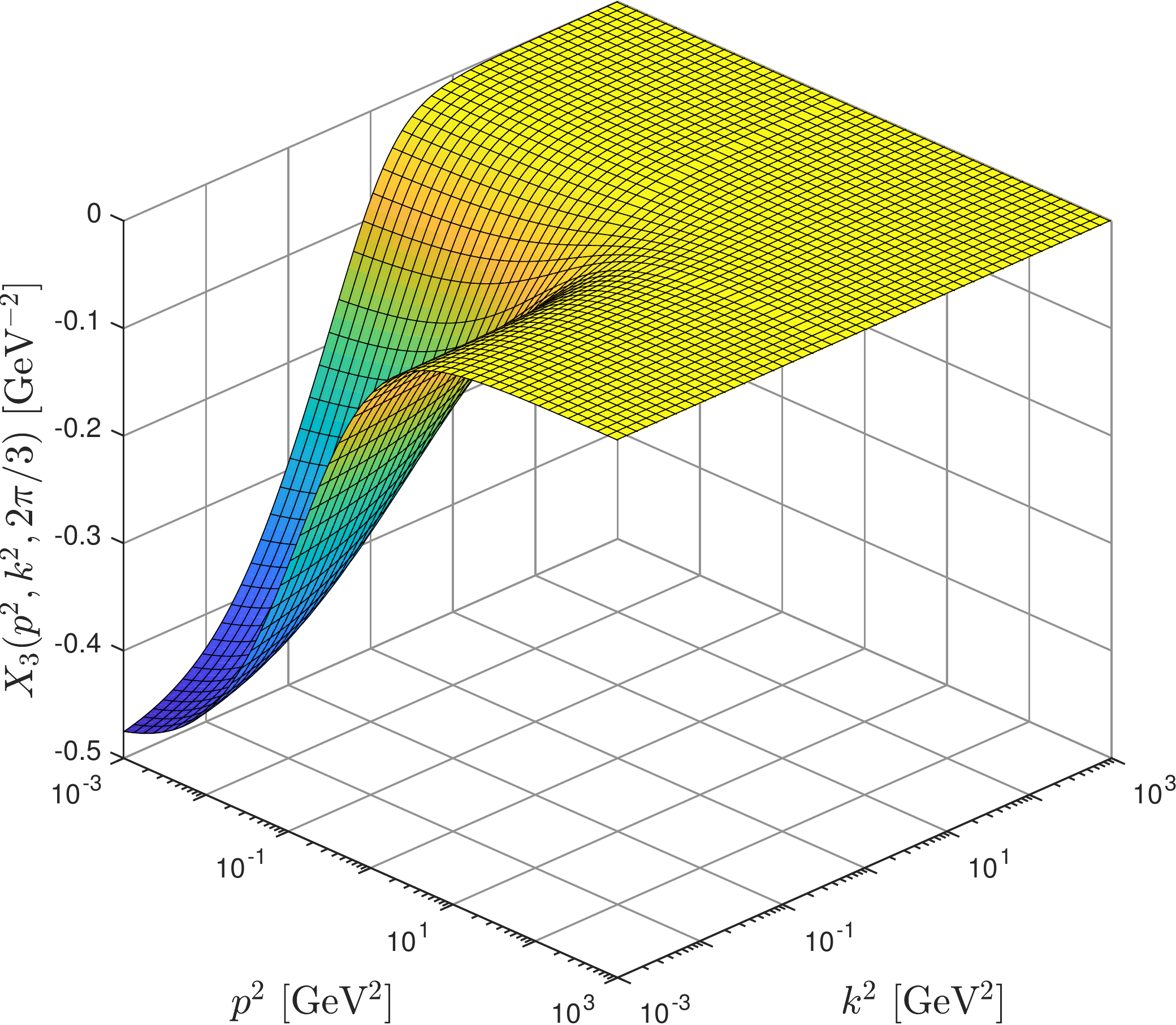}
\end{minipage}
\caption{\label{fig:Xi}  The form factors $X_i(p^2,k^2,2\pi/3)$,
obtained as solution of the coupled 
system given by Eqs.~\eqref{gAB1} and~\eqref{generalx}, when $\alpha_s=0.28$  and 
$\theta=2\pi/3$.}
\end{figure}

\begin{figure}[!hb]
\begin{minipage}[b]{0.40\linewidth}
\centering
\includegraphics[scale=0.37]{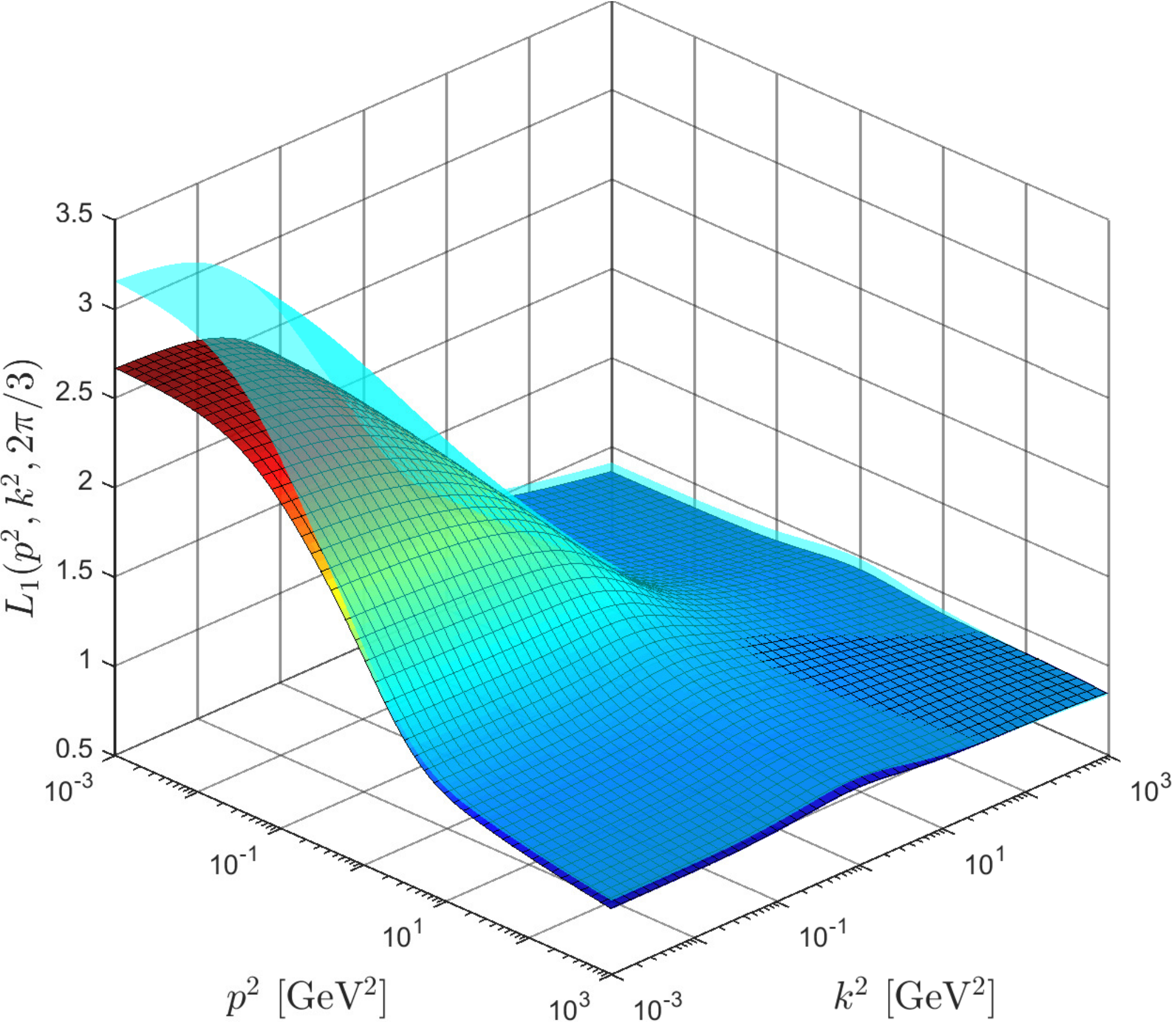}
\end{minipage}
\hspace{1.0cm}
\begin{minipage}[b]{0.45\linewidth}
\includegraphics[scale=0.37]{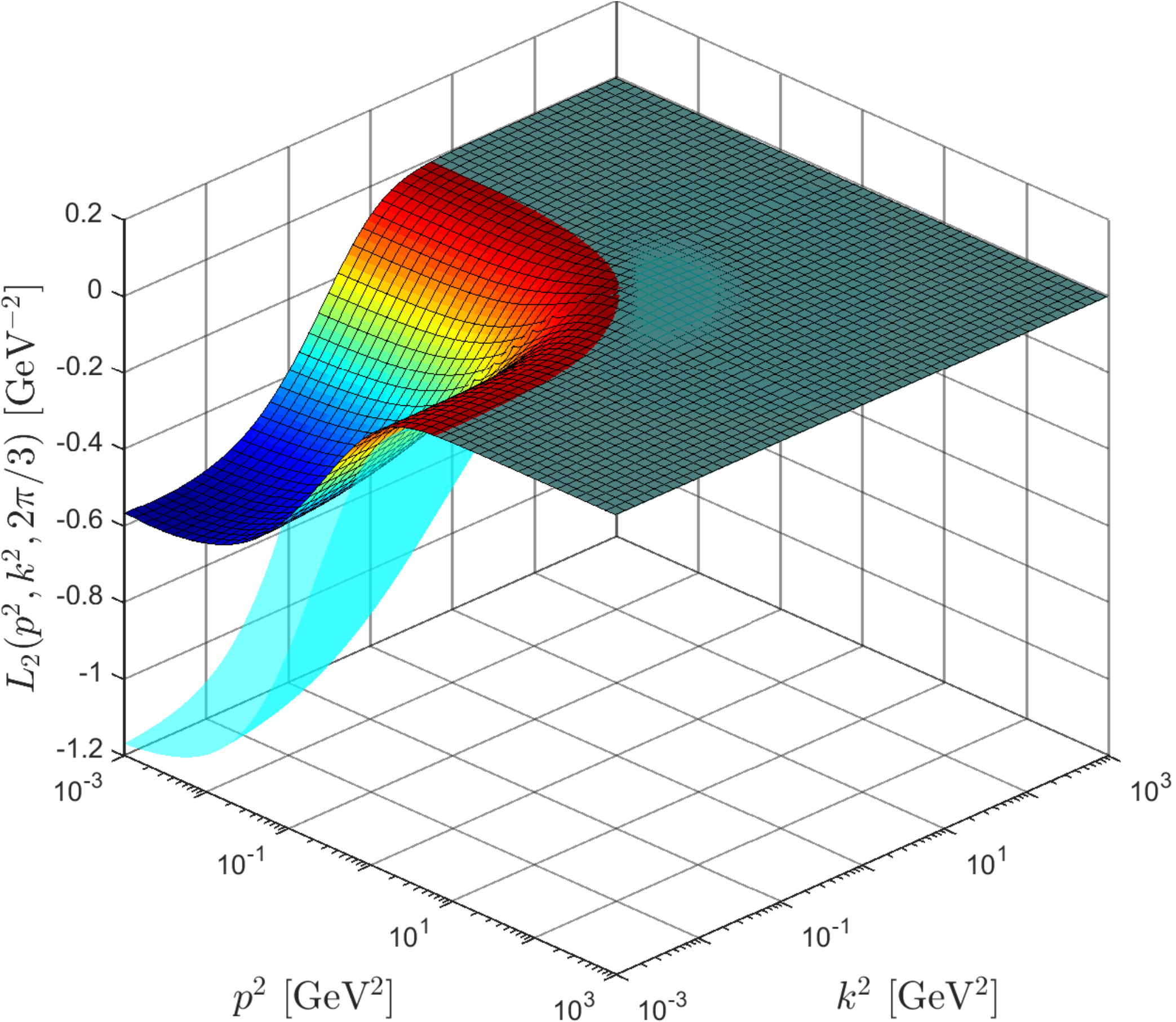}
\end{minipage}
\begin{minipage}[b]{0.40\linewidth}
\centering
\includegraphics[scale=0.37]{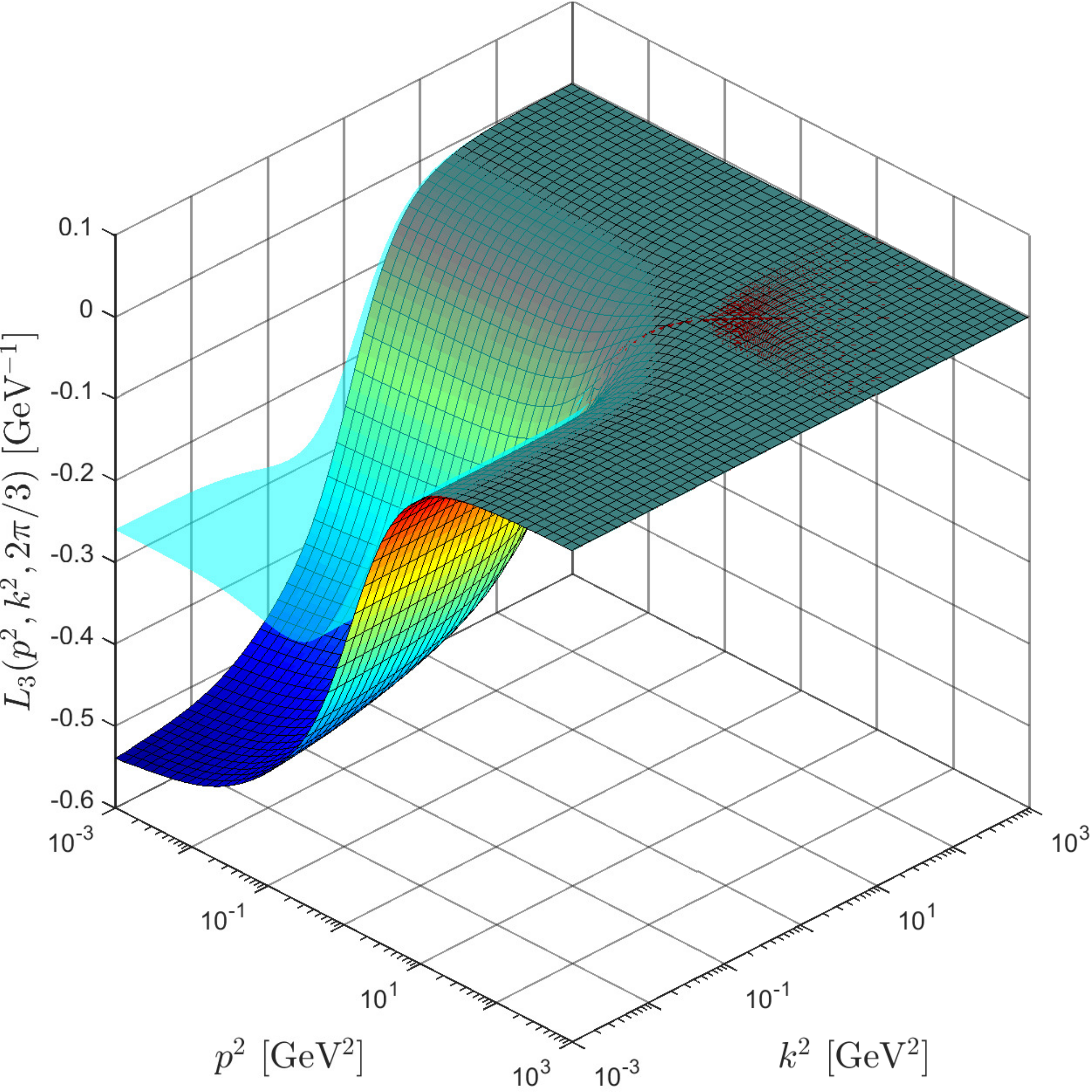}
\end{minipage}
\hspace{1.0cm}
\begin{minipage}[b]{0.45\linewidth}
\includegraphics[scale=0.36]{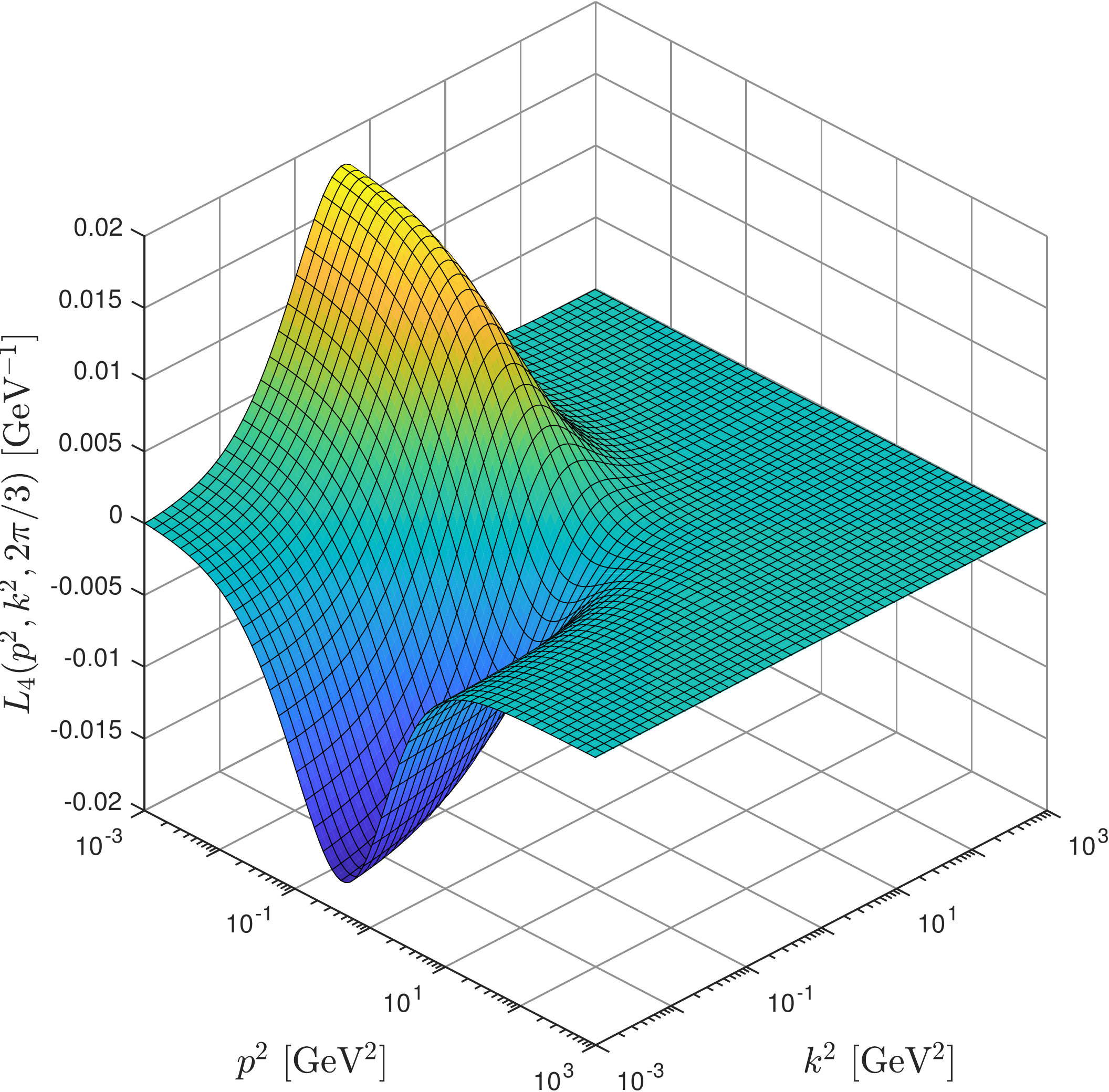}
\end{minipage}
\caption{\label{fig:LSDE}  The  quark-gluon form factors $L_i$
obtained by substituting into  Eq.~\eqref{expLi} the solutions of the coupled 
system given by Eqs.~\eqref{gAB1} and~\eqref{generalx}. The results
represent the case where $\alpha_s=0.28$  and 
$\theta=2\pi/3$.}
\end{figure}

In order to appreciate how  ${\mathcal M}(p)$ and $A^{-1}(p)$ are affected 
by the inclusion of $H$ (or, equivalently, the $X_i$) in the construction of the quark-gluon vertex,
in Fig.~\ref{mass_1G} we compare  
the solutions obtained with the full $\Gamma^{\STI}$ 
(blue continuous curves) with those computed using the $\Gamma^{\FBC}_{\mu}$ of Eq.~\eqref{FBC} (orange dashed lines). 
Evidently, the former solutions produce
higher ${\mathcal M}(p)$ compared to the latter,
in the entire range of momentum.
Of course, the quantitative
difference between ${\mathcal M}(p)$ and  ${\mathcal M}_{\FBC}(p)$  depends on  the precise value of $\alpha_s$: 
smaller values for $\alpha_s$ increase the difference between  ${\mathcal M}(p)$ and ${\mathcal M}_{\FBC}(p)$.
In particular, within the range of momenta between \mbox{$[0,780\,\mbox{MeV}]$}, we observe a difference of approximately $32\,\%$, $21\,\%$, and $19\,\%$, when $\alpha_s=0.24$, $\alpha_s=0.28$, and $\alpha_s=0.30$, respectively. A similar pattern is observed  in the  results for  $A^{-1}(p)$ and $A^{-1}_{\FBC}(p)$.

The remaining  four quantities are shown in the 3D plots of Fig.~\ref{fig:Xi}. Specifically,
we present a typical  set of results for 
the form factors $X_i$, where 
$\alpha_s=0.28$  and $\theta=2\pi/3$. 
We notice that all $X_i(p^2,k^2,2\pi/3)$ are infrared finite. In addition, 
all curves tend asymptotically to  their expected perturbative behaviors.

The $X_i$ computed in the previous step are subsequently fed into the Euclidean version of Eq.~\eqref{expLi}, 
thus furnishing the corresponding form factors $L_i$, shown in Fig.~\eqref{fig:LSDE}, where, as before,
$\alpha_s=0.28$ and  $\theta=2\pi/3$.

 As we can see, the behavior of the $L_i$ (colorful surfaces) is rather similar to that obtained in Ref.~\cite{Aguilar:2016lbe},  where  $A(p)$ and $B(p)$ were treated as ``external'' quantities. 
 As discussed in that work, the properties
 of the $L_i$ may be summarized as follows:  {\it(i)} the four form factors  are infrared finite in the entire range of momenta; {\it(ii)} the $L_i$ obtained indicate considerable deviations from  the $L_i^{\FBC}$ represented by the cyan surface, given by Eq.~\eqref{FBC}; {\it(iii)}  although $L_4$ is a non-vanishing quantity, its size is considerably
 suppressed  for all momenta,
 and {\it(iv)}  $L_2$ displays the most pronounced changes,
 because it is particularly sensitive to the details of the shape of $A(p)$. 

%
\begin{figure}[!h]
\begin{minipage}[b]{0.45\linewidth}
\centering
\includegraphics[scale=0.35]{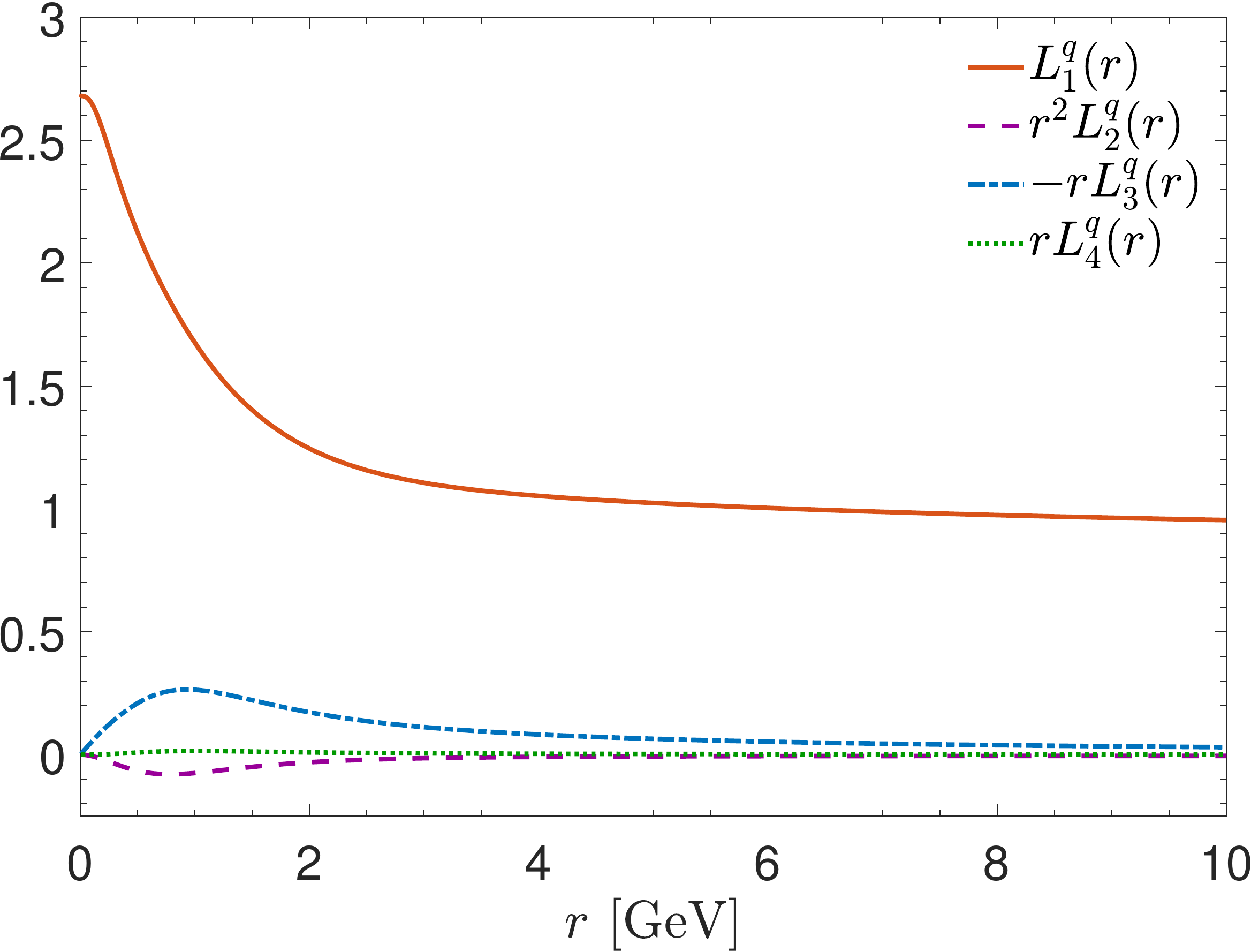}
\end{minipage}
\hspace{1.0cm}
\begin{minipage}[b]{0.45\linewidth}
\includegraphics[scale=0.35]{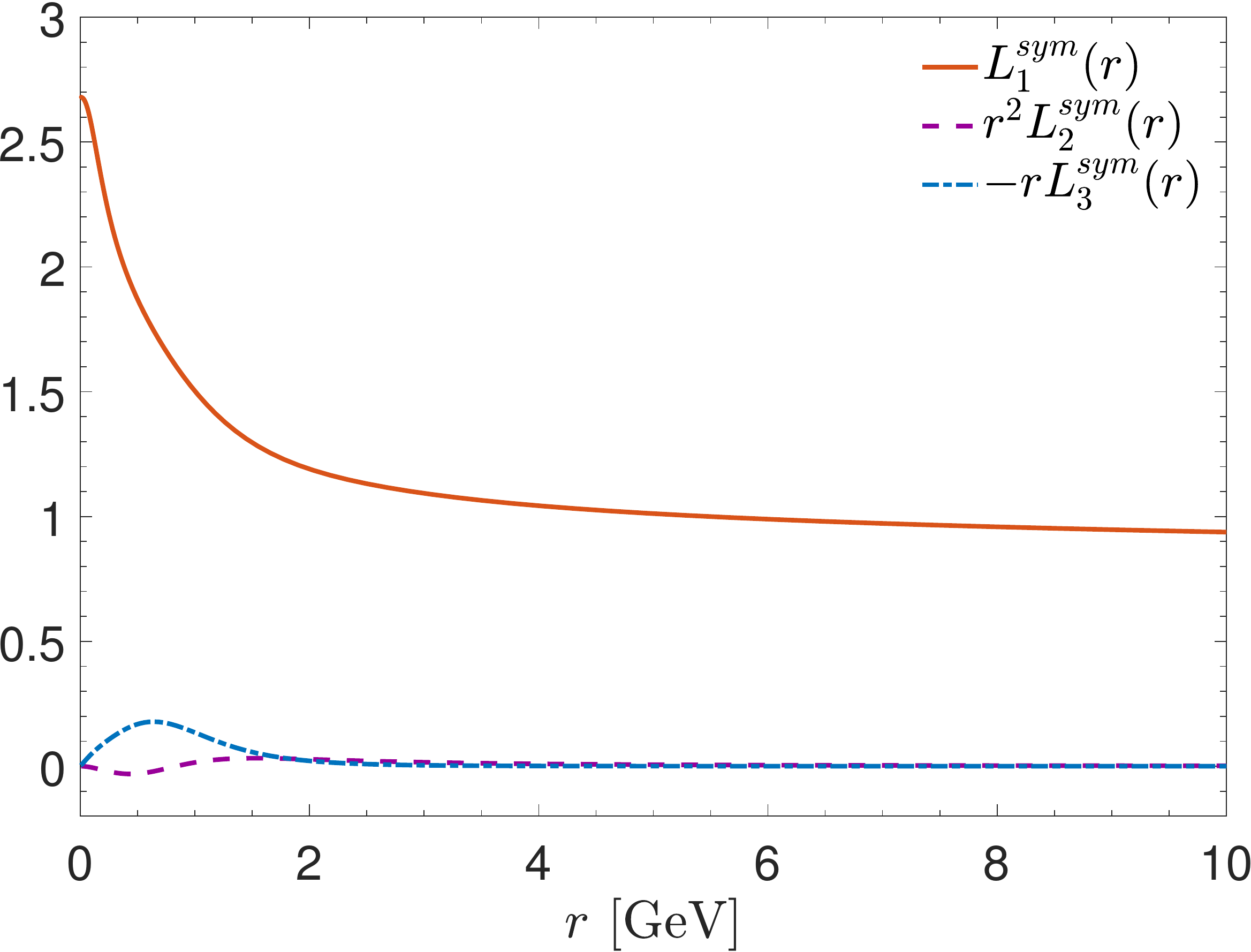}
\end{minipage}
\caption{\label{kinematic} The form factors $L_i$ for
different kinematic configurations. The $L_i^{q}(r)$ (left panel) 
correspond to the soft quark configuration, while the $L_i^{sym}(r)$ (right panel)
to the totally symmetric configuration.}
\end{figure}
 
Given that we have derived $L_i$ for general configurations,
we may easily single out  two special kinematics cases, namely 
{\it(i)} the ``soft quark'' limit, obtained as $p\to 0$, 
and  {\it(ii)} the ``totally symmetric'' limit,
where $p^2=k^2=q^2$ and $\theta=2\pi/3$.
Evidently, in these limits the $L_i$ become functions of  a single momentum, to be indicated by $r$; 
we will denote the corresponding form factors by $L_i^{q}(r)$ and $L_i^{sym}(r)$, respectively.  
In Fig.~\ref{kinematic} we show the corresponding results, with 
$L_i^{q}(r)$ on the left panel, and $L_i^{sym}(r)$ on the right. 
Note that, at the level of the  3D plots shown in Fig.~\ref{fig:LSDE}, 
the $L_i^{sym}(r)$ correspond to 
the ``slices'' defined by the planes $p=k$,  where $\theta=2\pi/3$. In particular, 
$L_4^{sym}(r)=0$. Moreover, in both cases,
we recover the expected perturbative behavior for large values of the momentum
(\mbox{$L_1=1$} and \mbox{$L_2=L_3=L_4=0$}).

It would be interesting to compare the above results with lattice simulations; however,  
the existing lattice data for the kinematic limits mentioned above are typically ``contaminated'' by 
contributions from $\Gamma^{\Tr}_{\mu}$~\cite{Skullerud:2004gp,Kizilersu:2006et}, due to an overall
contraction by $P_{\mu\nu}(q)$ (in the Landau gauge). 
For the case of the ``soft-gluon'' configuration, $q\to 0$, 
a detailed comparison both with the lattice  and with results
found with different functional approaches  
has been performed in~\cite{Aguilar:2016lbe}. Since the present results and those
of~\cite{Aguilar:2016lbe} are quite similar, a further comparison is of limited 
usefulness and will be omitted from the present work. 

\begin{figure}[!h]
\begin{minipage}[b]{0.45\linewidth}
\centering
\includegraphics[scale=0.35]{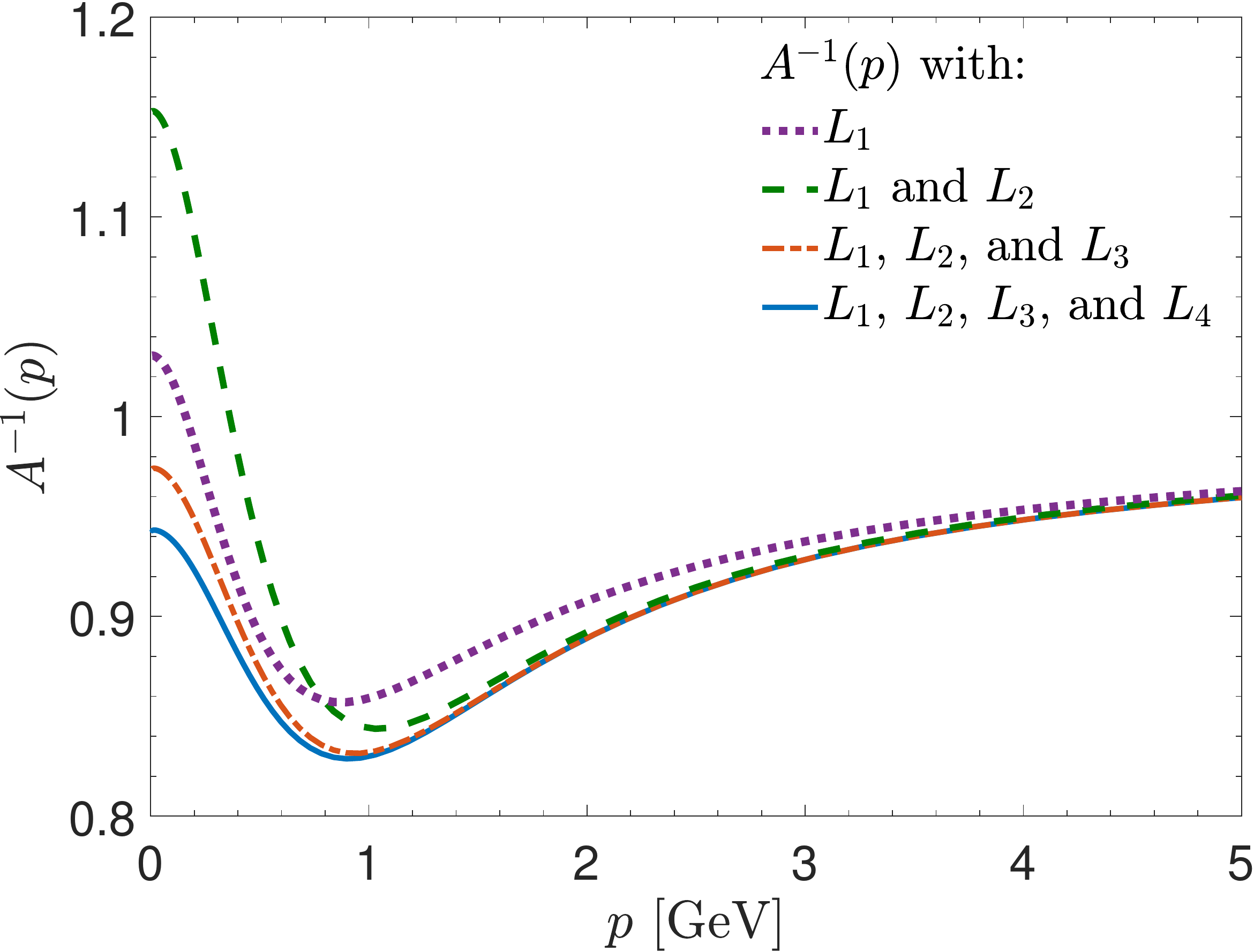}
\end{minipage}
\hspace{1.0cm}
\begin{minipage}[b]{0.45\linewidth}
\includegraphics[scale=0.35]{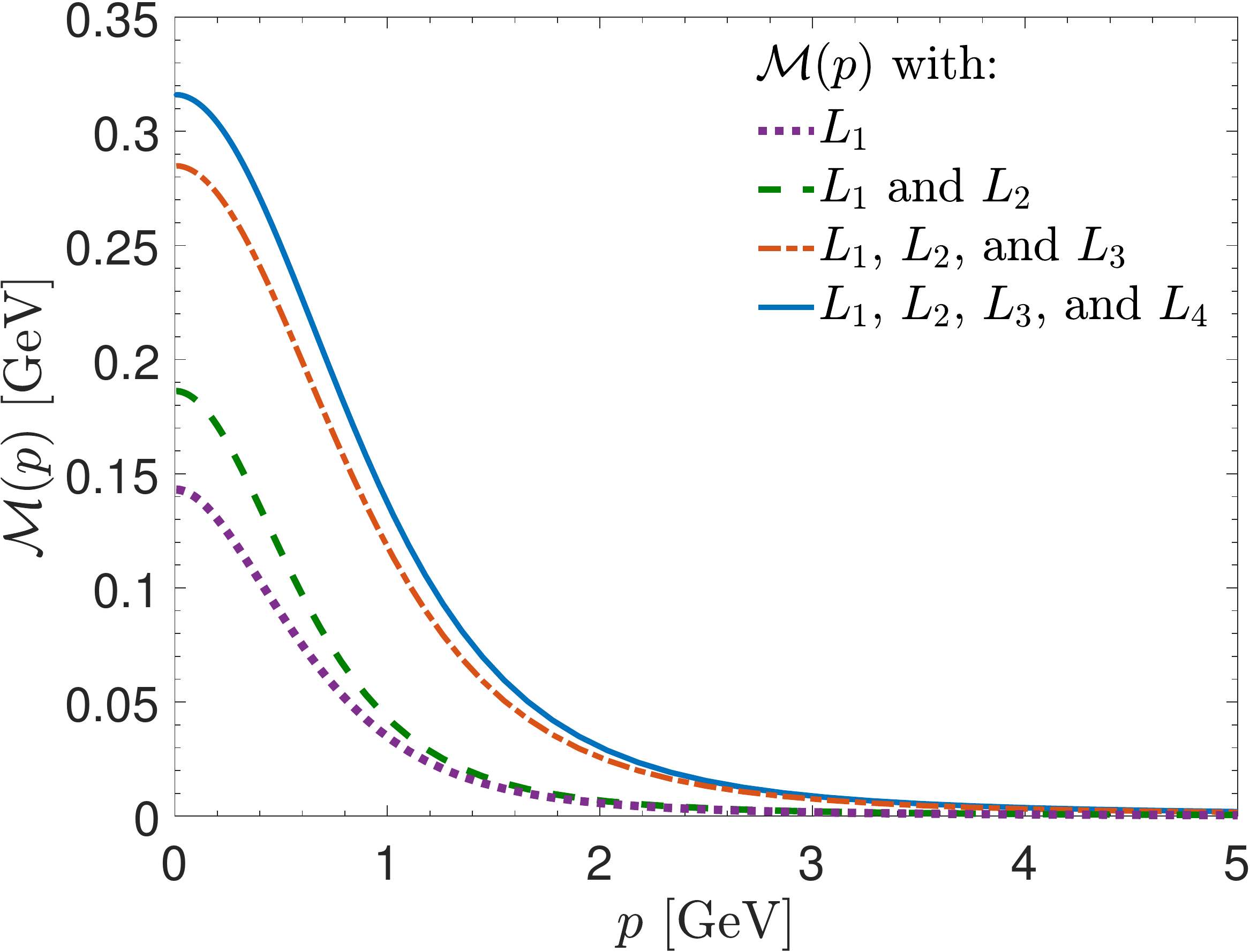}
\end{minipage}
\caption{\label{steps} Individual contributions
of the quark-gluon form factors $L_i$ on the {\it (i)} quark wave function, $A^{-1}(p)$ (left panel), and {\it (ii)} dynamical quark mass, ${\mathcal M}(p)$ (right panel).}
\end{figure}

Next, we  turn our attention to the numerical impact of each individual $L_i$ on the value of the 
dynamical quark mass. 
The results of this exercise are presented in Fig.~\ref{steps}, where in both panels
we show the corresponding $A^{-1}(p)$ and ${\mathcal M}(p)$, which are generated  
as we turn on, one by one, the form factors $L_i$ that compose
the kernels  ${\mathcal K}_{\rm{\s A}}$ and ${\mathcal K}_{\rm{\s B}}$, given by Eqs.~\eqref{kernelAB}.
Clearly, the largest numerical contribution comes from $L_1$, which
is responsible for generating $54\%$ of  ${\mathcal M}(0)$.
In addition, $L_2$ furnishes $13\%$ of the ${\mathcal M}(0)$ value, while  
$L_3$ contributes another $23\%$. Particularly interesting  is the impact of $L_4$; even
though it is rather suppressed [see Fig.~\ref{fig:LSDE}], 
and is usually neglected in related studies~\cite{Aguilar:2010cn,Fischer:2003rp,Roberts:1994dr},
$L_4$ provides, rather  unexpectedly,  $10\%$ of  ${\mathcal M}(0)$.   
  
\subsection{\label{Num_C}Varying the form of ${\mathcal C}(q)$}

In order to determine the influence of the functions ${\mathcal C}_i(q)$ 
on the coupled system, we repeat the analysis using ${\mathcal C}_2(q)$ and ${\mathcal C}_3(q)$ instead of ${\mathcal C}_1(q)$
[equivalently, \mbox{${\mathcal R}_1(q) \to {\mathcal R}_2(q)$}, or
\mbox{${\mathcal R}_1(q) \to {\mathcal R}_3(q)$}].

\begin{figure}[!h]
\begin{minipage}[b]{0.3\linewidth}
\centering
\includegraphics[scale=0.22]{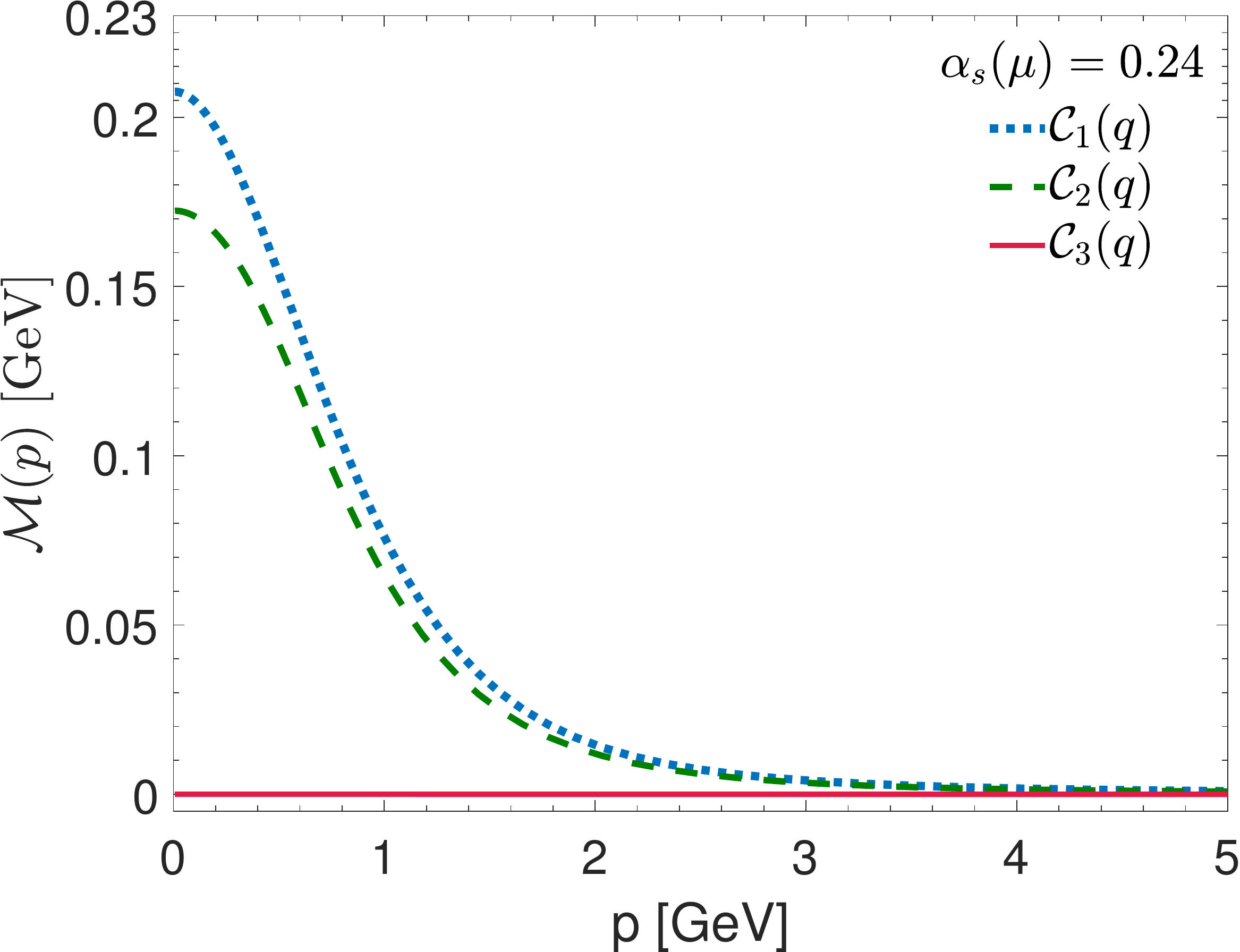}
\end{minipage}
\hspace{0.3cm}
\begin{minipage}[b]{0.3\linewidth}
\includegraphics[scale=0.22]{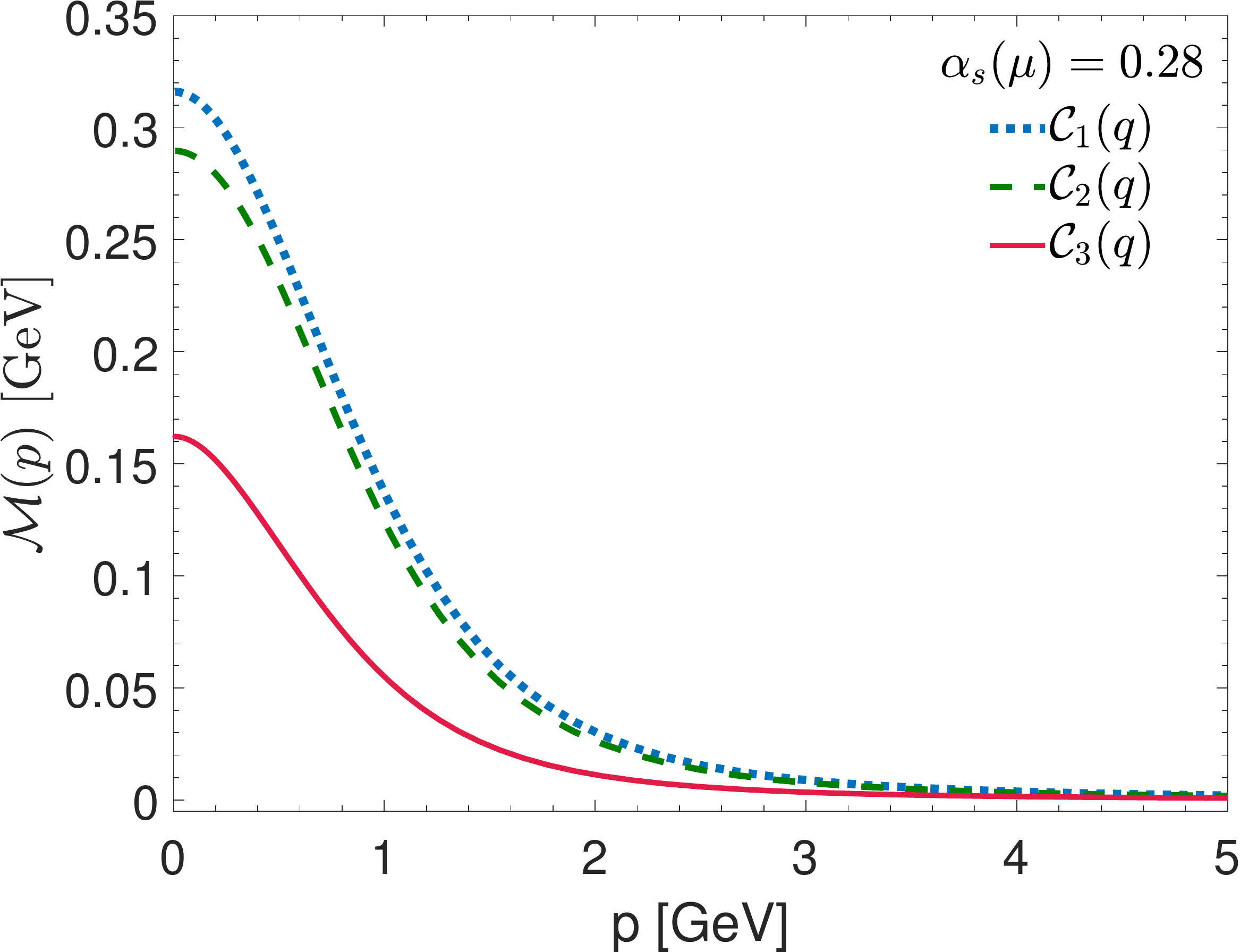}
\end{minipage}
\hspace{0.3cm}
\begin{minipage}[b]{0.3\linewidth}
\includegraphics[scale=0.22]{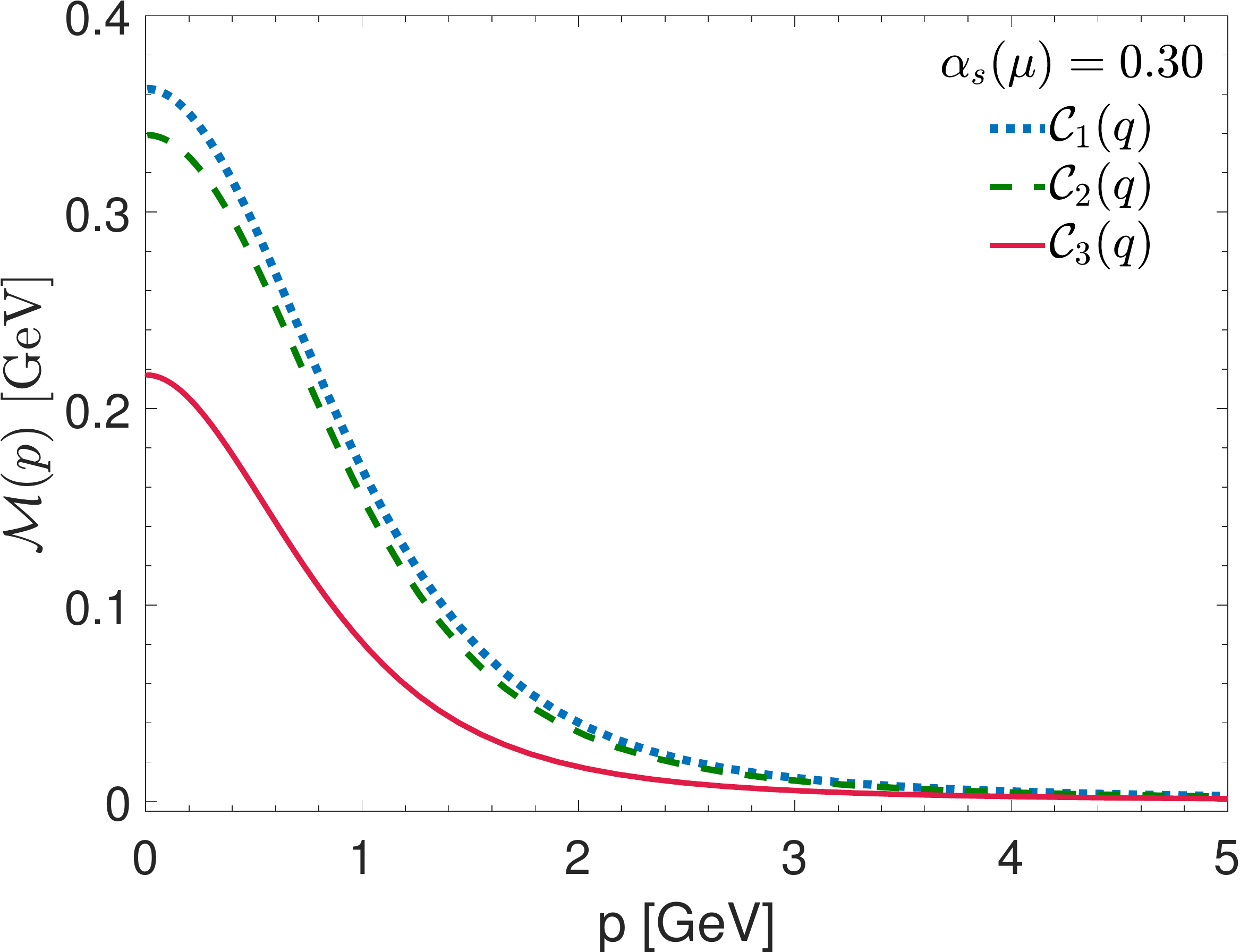}
\end{minipage}
\begin{minipage}[b]{0.3\linewidth}
\centering
\includegraphics[scale=0.22]{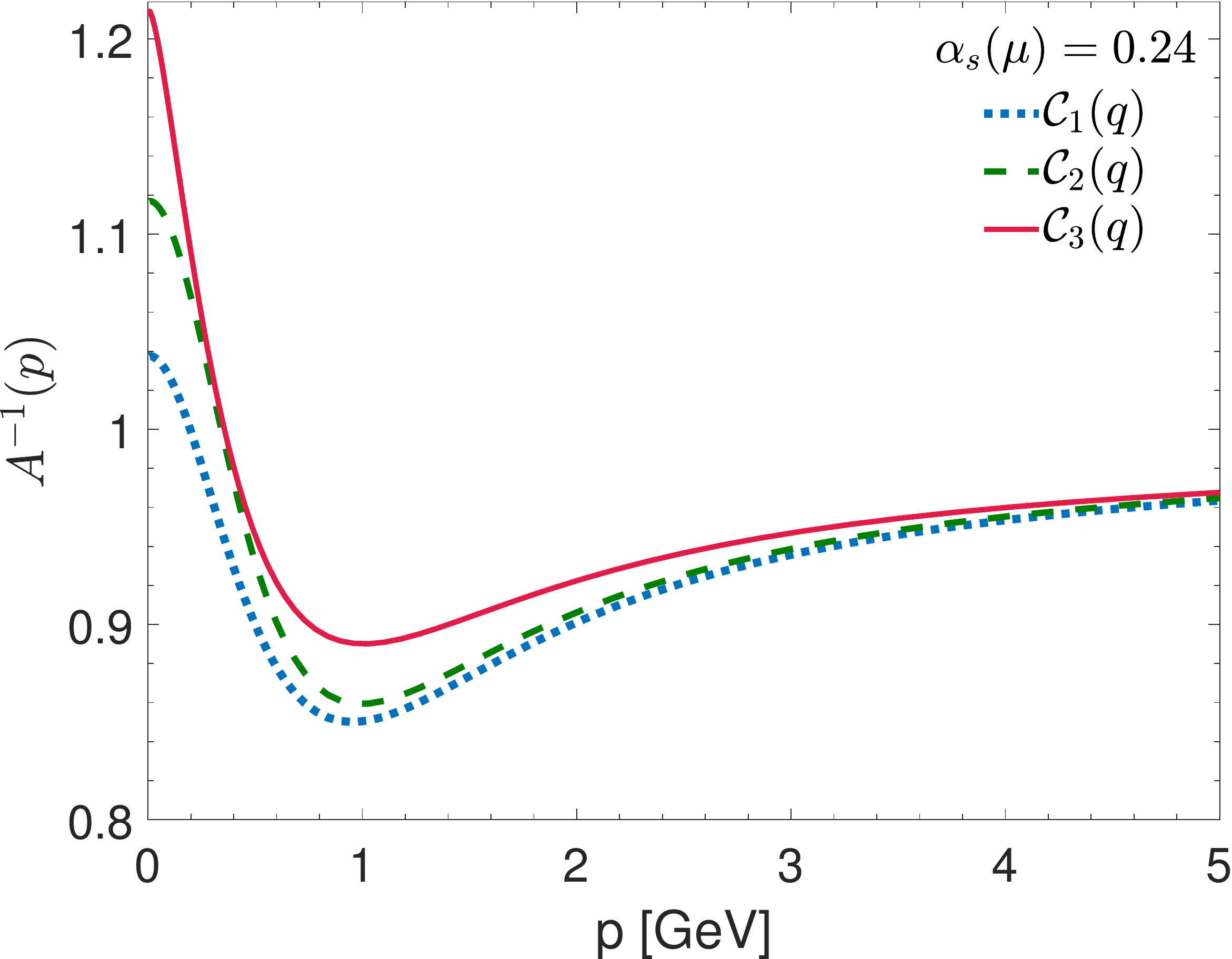}
\end{minipage}
\hspace{0.3cm}
\begin{minipage}[b]{0.3\linewidth}
\includegraphics[scale=0.22]{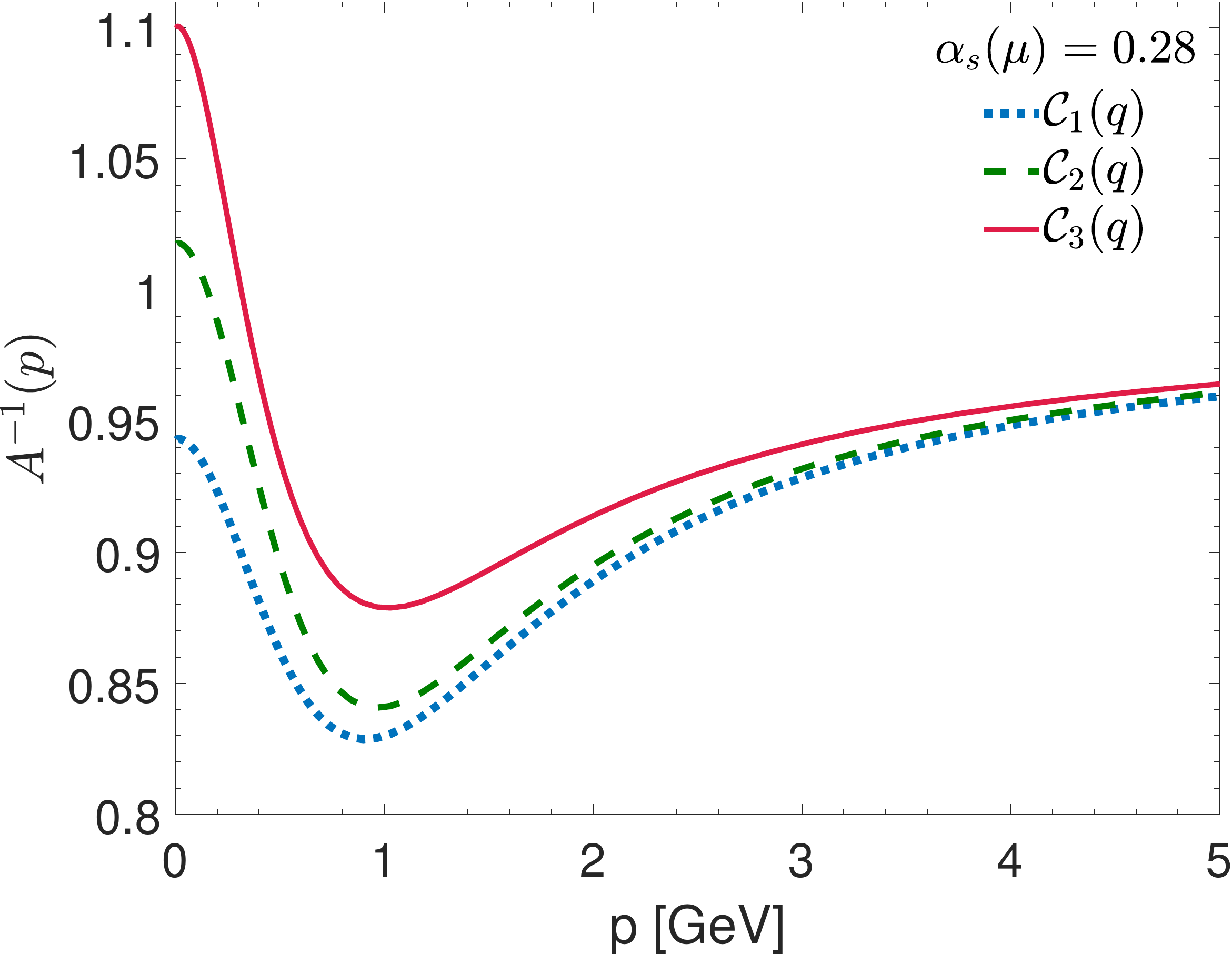}
\end{minipage}
\hspace{0.3cm}
\begin{minipage}[b]{0.3\linewidth}
\includegraphics[scale=0.22]{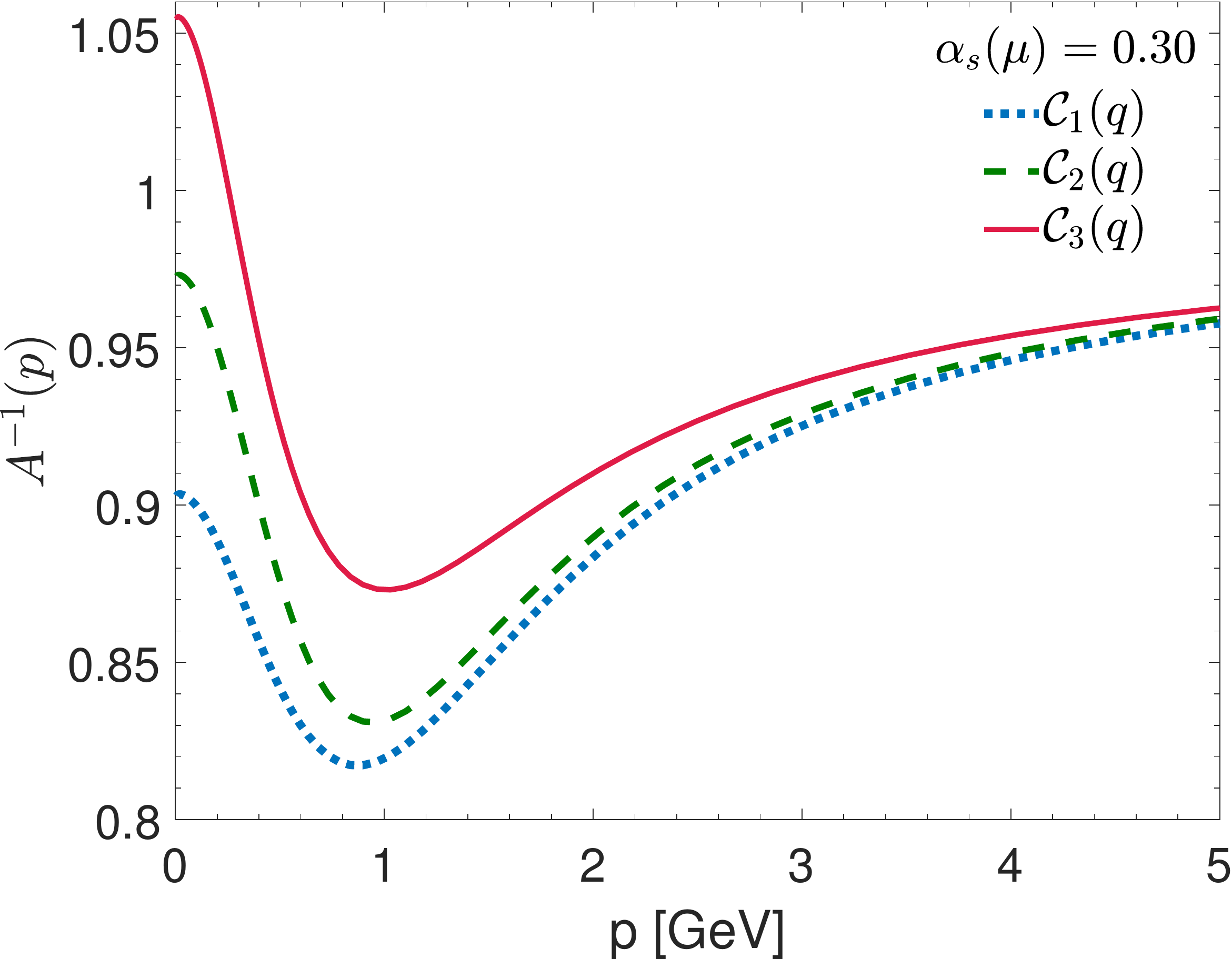}
\end{minipage}
\vspace{-0.3cm}
\caption{\label{fig:Hansatze} Comparison of the  dynamical quark masses,~${\mathcal M}(p)$, (top panels) and the quark wave function, $A^{-1}(p)$, (bottom panels) obtained when we employ the three  {\it Ans\"atze} for  ${\mathcal C}_i(q)$ given by Eq.~\eqref{rfunc} for different values of $\alpha_s$.}
\end{figure}

In Fig.~\eqref{fig:Hansatze} we perform a comparative analysis of the  
$A^{-1}(p)$ and ${\mathcal M}(p)$ obtained when
we employ the three  {\it Ans\"atze} for  ${\mathcal C}_i(q)$, given by Eq.~\eqref{rfunc},
for different values of $\alpha_s$.

Although ${\mathcal C}_2(q)$ is significantly more
suppressed in the deep infrared compared to ${\mathcal C}_1(q)$ and ${\mathcal C}_3(q)$ [see Fig.~\ref{func_ren}], one can observe that, essentially, the first two models 
generate quark masses of comparable size: the masses obtained using  
${\mathcal C}_1(q)$ (blue dotted curve) are  slightly larger than those
coming from ${\mathcal C}_2(q)$ (green dashed curve).
Clearly, the difference in the results 
obtained with ${\mathcal C}_1(q)$ and ${\mathcal C}_2(q)$
decreases as $\alpha_s$ increases; in particular, 
the difference between the corresponding  ${\mathcal M}(0)$
computed with $\alpha_s=0.24$, $\alpha_s=0.28$, and $\alpha_s=0.30$
is about $20\%$, $10\%$, and $6\%$, respectively.

Instead, ${\mathcal C}_3(q)$ does not provide sufficient strength 
to the kernel of the gap equation \eqref{gAB1} to trigger the onset of
the dynamical mass generation, when $\alpha_s=0.24$ (red continuous curve in the top left panel). Although for higher values of $\alpha_s$ the chiral symmetry
is eventually broken,  one 
notices that the values of masses obtained are phenomenologically disfavored;
specifically,
one finds \mbox{$160$ MeV} for $\alpha_s=0.28$, and \mbox{$217$ MeV} when  
$\alpha_s=0.30$.

We emphasize that the mass pattern emerging from the above exercise is consistent with what one would
expect on general grounds. Indeed, as is well-established by now, 
the  support of the gap equation kernel in the intermediate region of
momenta is crucial for the generation of  phenomenologically compatible quark masses~\cite{Fischer:2003rp,Aguilar:2010cn},
while modifications of that kernel in the  deep infrared
do not affect significantly the resulting quark mass~\cite{Roberts:1994dr,Maris:1999nt}.
Consequently, the origin of the small difference in the  ${\mathcal M}(p)$ obtained with the first two models  
can be naturally attributed to the slight
suppression that  ${\mathcal C}_2(q)$ displays in the region of \mbox{[$1-2]$ GeV}
in comparison with ${\mathcal C}_1(q)$, whereas 
the sizable suppression of ${\mathcal C}_3(q)$
in the range of \mbox{[$0.5-1.5$] GeV} prohibits or reduces substantially the generation of a quark mass.

We conclude this subsection by presenting in Table~\ref{table_impact}  a detailed analysis of the impact of the scattering kernel $H$ on the dynamical mass generation,  as we vary the function ${\mathcal C}_i(q)$. We will restrict ourselves to the
comparison of  the  values for ${\mathcal M}_{\FBC}(0)$ and ${\mathcal M}(0)$; we remind the reader that,
in the former case, $H$ assumes its tree-level value, while the latter  
is obtained from solving the system.
The  impact will be quantified through the 
 relative percentage difference 
  \mbox{$I_{\rm H}=[{\mathcal M}(0)/{\mathcal M}_{\FBC}(0) -1]\times 100\% $}.
Independently of the form that  ${\mathcal C}_i(q)$ assumes,  one notices that $I_{\rm H}$
depends on  the value of $\alpha_s$, reaching larger values as $\alpha_s$ decreases. 
Interestingly enough, as we reach
phenomenologically relevant 
values for ${\mathcal M}(0)$ (\ie in the range \mbox{$280-360$ MeV}), 
$I_{\rm H}$  practically stabilizes around $20\%$.

\vspace{0.5cm}
\begin{table}[h]
\begin{tabular}{|c|c|c|c|c|c|c|c|c|c|}
\hline
\hline
  &\multicolumn{3}{c|}{ Masses with ${\mathcal C}_1(q)$ [MeV] } &\multicolumn{3}{c|}{ Masses with ${\mathcal C}_2(q)$ [MeV]}
  &\multicolumn{3}{c|}{Masses with ${\mathcal C}_3(q)$ [MeV] }\\ 
  \hline
 \cline{2-4} \cline{5-7}\cline{8-10}
 $\alpha_s$ &\;${\mathcal M}_{\FBC}(0)$\;&${\mathcal M}(0)$ & $I_{\rm H}$ &${\mathcal M}_{\FBC}(0)$& ${\mathcal M}(0)$ & $I_{\rm H}$&${\mathcal M}_{\FBC}(0)$&${\mathcal M}(0)$ &$I_{\rm H}$\\ 
\colrule
0.24&157&207&$32\%$&114&172&$51\%$&0&0&$0\%$\\
0.28&261&316&$21\%$&231&286&$24\%$&86&162&$88\%$\\ 
0.30&305&362&$19\%$&278&339&$22\%$&142&217&$53\%$ \\
\hline
\hline
 \end{tabular}
  \vspace{0.25cm}
\caption{Comparison of the values obtained for ${\mathcal M}_{\FBC}(0)$ and ${\mathcal M}(0)$  when we employ the three {\it Ans\"atze}  ${\mathcal C}_i(q)$ of  Eq.~\eqref{rfunc}.}
 \label{table_impact}    
 \end{table}

\subsection{\label{Num_fit} Fits for the constituent quark mass}

It turns out that all running quark masses  ${\mathcal M}(p)$
presented in the Fig.~\ref{fig:Hansatze}  may be accurately fitted by the following physically motivated fit
\bea
{\mathcal M}(p) = \frac{{\mathcal M}_1^3}{{\mathcal M}_2^2 + p^2\left[\ln(p^2+{\mathcal M}_3^2)/{\Lambda^2}\right]^{1-\gamma_f}} \,,
\label{fit2}
\eea
where ~\mbox{$({\mathcal M}_1,{\mathcal M}_2,{\mathcal M}_3)$} are the three adjustable ``mass'' parameters,
and  \mbox{$\Lambda=270\,\mbox{MeV}$}. 

The above formula constitutes a simple infrared completion of Eq.~\eqref{fituv}, where 
the  presence of the ${\mathcal M}_2$ in the denominator enforces the saturation of ${\mathcal M}(p)$
at the origin, while the ${\mathcal M}_3$  in the argument of the logarithm 
 improves the convergence of the fitting procedure. 

 It turns out that the expression 
\bea
{\mathcal M}(p) = \frac{{\mathcal M}_0}{1+\left(p^2/\lambda^2\right)^{1+d}}\,,
\label{fit}
\eea
is yet another excellent fit for  all our results for ${\mathcal M}(p)$.
The functional form of \1eq{fit}
may be easier to handle when numerical integrations
of ${\mathcal M}(p)$ are involved.

\begin{figure}[!t]
\begin{minipage}[b]{0.45\linewidth}
\centering
\includegraphics[scale=0.35]{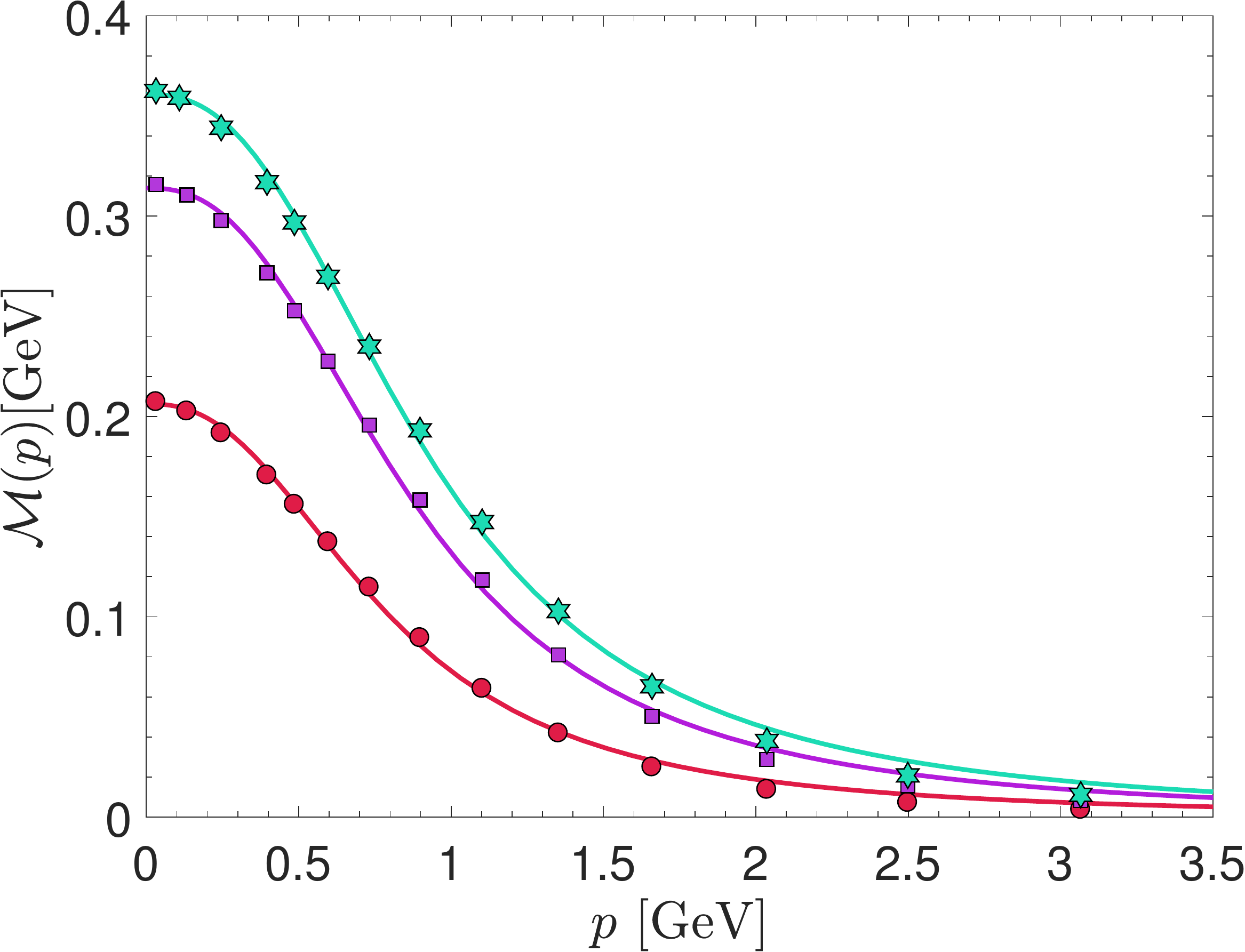}
\end{minipage}
\hspace{1.0cm}
\begin{minipage}[b]{0.45\linewidth}
\includegraphics[scale=0.35]{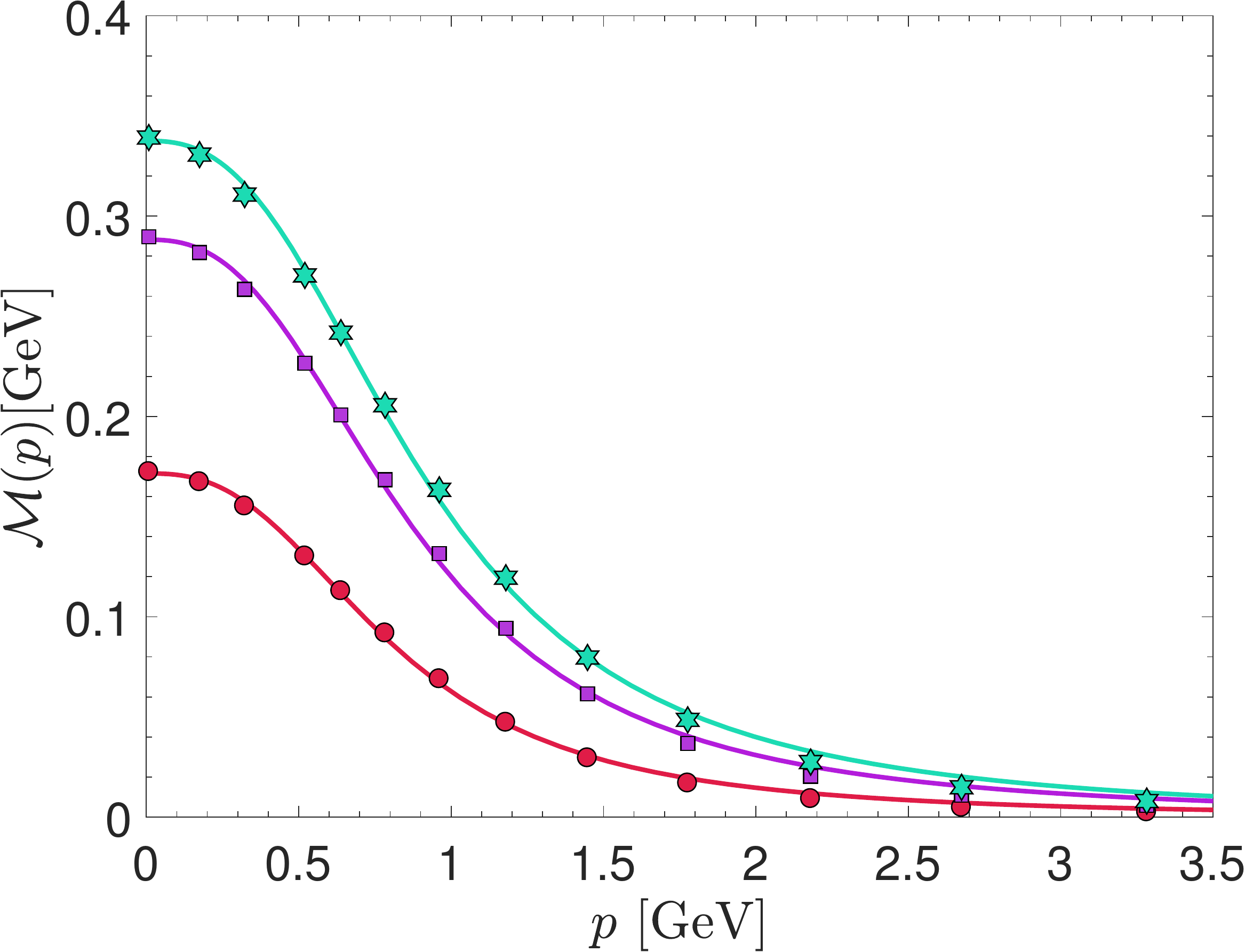}
\end{minipage}
\caption{\label{figure:fit} The numerical solution
for  ${\mathcal M}(p)$ obtained using the ${\mathcal C}_1(q)$ (left panel)
and ${\mathcal C}_2(q)$ (right panel). In each panel we display the solutions
for \mbox{$\alpha_s=0.24$} (red circles),  \mbox{$\alpha_s=0.28$} (purple squares) and  \mbox{$\alpha_s=0.30$} (green stars). The continuous curves
represent the fit of Eq.~\eqref{fit}.}
\end{figure}

In Fig.~\ref{figure:fit} we superimpose the numerical 
 solutions when \mbox{$\alpha_s=0.24$} (red circles),  \mbox{$\alpha_s=0.28$} (purple squares), and  \mbox{$\alpha_s=0.30$} (green stars) for ${\mathcal C}_1(q)$ (left panel) and ${\mathcal C}_2(q)$  (right panel) and the fit of Eq.~\eqref{fit} (continuous curves). Since it is not possible
to notice any  sizable quantitative difference between the fits produced either with Eqs.~\eqref{fit2} or~\eqref{fit},  
in Fig.~\ref{figure:fit} we only  show the curves for Eq.~\eqref{fit}. 
The corresponding sets of parameters \mbox{$({\mathcal M}_1,{\mathcal M}_2,{\mathcal M}_3)$}
and $({\mathcal M}_0,\lambda, d)$ are quoted in the Table~\ref{table_fit}.  All fits have a reduced \mbox{$\chi^2=0.99$}.

\begin{table}[h]
\begin{tabular}{|c|c|c|c|c|c|c|}
\hline
\hline
  &\multicolumn{3}{c|}{ Fit given by Eq.~\eqref{fit2} } &\multicolumn{3}{c|}{ Fit given by Eq.~\eqref{fit}}\\ 
  \hline
 \cline{2-4} \cline{5-7}
 $\alpha_s$  &${\mathcal M}_1$ [MeV]& ${\mathcal M}_2$ [GeV] & ${\mathcal M}_3$ [MeV] &\;${\mathcal M}_0$ [MeV]\;&$\lambda$ [MeV] &  $\qquad d \qquad $\\ 
\colrule
0.24 with ${\mathcal C}_1(q)$&601&1.03&404&206&780&0.22\\ 
0.24 with ${\mathcal C}_2(q)$&572&1.04&270&171&809&0.31\\
0.28 with ${\mathcal C}_1(q)$&758&1.18&426&314&878&0.25\\ 
0.28 with ${\mathcal C}_2(q)$&715&1.12&270&288&876&0.28\\ 
0.30 with ${\mathcal C}_1(q)$&824&1.25&358&361&925&0.25 \\
0.30 with ${\mathcal C}_2(q)$&772&1.16&270&337&914&0.28 \\
\hline
\hline
 \end{tabular}
  \vspace{0.25cm}
\caption{The sets of adjustable parameters employed for the fits given by 
Eqs.~\eqref{fit2} and~\eqref{fit}. All fits have a reduced \mbox{$\chi^2=0.99$}.}
 \label{table_fit}   
 \end{table}

\subsection{\label{Num_fpi}Estimating the pion decay constant}

In order to appreciate the impact of  $H$ on a physical observable
sensitive to the dynamical quark mass, we turn to the pion decay constant, $f_{\pi}$.
For its computation we use 
an improved version  of the  Pagels-Stokar-Cornwall formula~\cite{Pagels:1979hd,Cornwall:1980zw} proposed in~\cite{Roberts:1994hh}, given by\footnote{The values of $f_{\pi}$ obtained from an 
  alternative expression given in Eq.~(6.27) of~\cite{Roberts:1994dr} are about $10\%$ lower.}. 
\be
f_{\pi}^2 = \frac{3}{8\pi^2}\int^{\infty}_0 \!\! dy
yB^2(y)\left\{
\sigma_{\!\s V}^2 -2\left[\sigma_{\!\s S}\sigma_{\!\s S}^{\prime} + y\sigma_{\!\s V}\sigma_{\!\s V}^{\prime}\right]
-y\left[\sigma_{\!\s S}\sigma_{\!\s S}^{\prime\prime} - (\sigma_{\!\s S}^{\prime})^2\right]\right. 
\left.-y^2\left[\sigma_{\!\s V}\sigma_{\!\s V}^{\prime\prime} - (\sigma_{\!\s V}^{\prime})^2\right]
\right\}\,,
\label{fpi}
\ee
where 
\bea
\sigma_{\!\s V} := \frac{A(y)}{yA^2(y)+B^2(y)}\,,
\qquad \qquad  \sigma_{\!\s S} := \frac{B(y)}{yA^2(y)+B^2(y)} \,.
\eea

The values quoted in the Table~\ref{table_fpi}  for $f_{\pi}$  should be compared to the experimental value \mbox{$f_{\pi}=93$ MeV}~\cite{Patrignani:2016xqp}. Evidently, ${\mathcal C}_3(q)$  produces the smallest set of values for $f_{\pi}$, since the corresponding ${\mathcal M}(p)$, entering in Eq.~\eqref{fpi}, are quite suppressed in comparison with the others
solutions obtained with   ${\mathcal C}_1(q)$ or  ${\mathcal C}_2(q)$. 
 Our analysis shows clearly a preference for $\alpha_s$ in the range of $0.28-0.30$, and
for the functional forms given by ${\mathcal C}_1(q)$ or  ${\mathcal C}_2(q)$. In addition,  one
notices that, for either ${\mathcal C}_1(q)$ or  ${\mathcal C}_2(q)$,
the relative percentage difference  between the values for $f_{\pi}$ obtained with $\Gamma^{\STI}$ and  $\Gamma^{\FBC}_{\mu}$ are approximately
$10\%$,  when $\alpha_s=0.28$ and $\alpha_s=0.30$. 
%
\begin{table}[!h]
\begin{tabular}{|c|c|c|c|c|c|c|}
\hline
\hline
  &\multicolumn{2}{c|}{$f_{\pi}$ with ${\mathcal C}_1(q)$} &\multicolumn{2}{c|}{$f_{\pi}$ with ${\mathcal C}_2(q)$ }&\multicolumn{2}{c|}{$f_{\pi}$ with ${\mathcal C}_3(q)$ }\\ 
  \hline
 \cline{2-3}   \cline{4-5} \cline{5-6}
 $\alpha_s$  &\;$\Gamma^{\FBC}_{\mu}$\;&$\Gamma^{\STI}_{\mu}$\;  &\;$\Gamma^{\FBC}_{\mu}$\; & \;$\Gamma^{\STI}_{\mu}$&\;$\Gamma^{\FBC}_{\mu}$\;&\;$\Gamma^{\STI}_{\mu}$\;\\  \colrule
 0.24&$62$&$73$& $52$&$67$&$0$& 0\\ 
 0.28&$87$&$97$&$83$&$93$&$40$&61\\ 
 0.30&$97$&$107$&$93$&$103$&$57$&75 \\
\hline
\hline
 \end{tabular}
 \caption{Values for $f_{\pi}$ computed with Eq.~\eqref{fpi} in [MeV]. 
 The six sets of results were calculated using the corresponding $A(p)$ and ${\mathcal M}(p)$ obtained with the three ${\mathcal C}_i(q)$ given by Eq.~{\eqref{rfunc}}, when we employ either
the ``minimal" non-abelian Ball-Chiu vertex, $\Gamma^{\FBC}_{\mu}$, or the 
complete $\Gamma^{\STI}$.}
 \label{table_fpi}    
 \end{table}

\vspace{-0.2cm}
\section{\label{sec:c}Discussion and Conclusions}
\vspace{-0.2cm}
In this article we have performed a detailed study of the dynamical quark mass pattern that emerges when the gap equation is coupled to the four dynamical equations that determine the structure of the quark-ghost kernel, $H$, and, in turn, the
STI-saturating part of the quark-gluon vertex, $\Gamma_{\mu}$. 
The analysis has been carried out in the Ball-Chiu tensorial basis,  
and the dynamical equations for  $H$ are derived  
within the one-loop dressed truncation scheme, under certain simplifying assumptions
for the vertices appearing in them. The corresponding gap equation that generates the
dynamical quark mass has been treated in the chiral limit (vanishing ``current'' mass).  

The numerical effect of including a non-trivial $H$ into the
construction of the $\Gamma_{\mu}$ that enters 
in the gap equation is rather sizable. Indeed, 
as we have seen in the Table~\ref{table_impact}, while its precise 
contribution depends on the value of $\alpha_s$, it accounts for approximately $20\%$ of the  dynamical quark mass generated, when ${\mathcal M}(0)$ is in the range of \mbox{$280-360$ MeV}.

The impact of $H$ on the dynamics of chiral symmetry breaking was also estimated indirectly,  
through the determination of the pion decay constant, $f_{\pi}$.
When  phenomenological compatible quark masses are generated,  we see that the 
inclusion of $H$ into $\Gamma_{\mu}$, \ie the transition $\Gamma^{\rm \s{BC}}_{\mu}\to \Gamma_{\mu}$,
amounts to a $10\%$ increase in the value of $f_{\pi}$. 

It is important to emphasize that in the present analysis a non-trivial 
 structure of the vertex form factor $L_4$ was included in the gap equation.
Despite the fact that $L_4$ is rather suppressed compared to $L_1$, $L_2$, $L_3$, 
as shown in 
the Fig.~\ref{fig:LSDE}, our findings indicate that it accounts for $10\%$ of total ${\mathcal M}(0)$ generated.
Therefore, $L_4$ contributes to the dynamical mass generation  practically with the same strength as $L_2$. 
This result, in turn, seems to suggest that $L_4$ provides 
a more ``focused'' support to the gap equation kernel, enhancing it  
precisely in the range of momenta that drive the onset of chiral symmetry breaking. 
To the best of our knowledge, such a concrete quantitative statement
on the impact of $L_4$ appears for the first time in the literature. 

Given that the multiplicative renormalizability of the quark propagator 
constitutes a notoriously difficult task,  
the restoration of the correct one-loop anomalous dimension for ${\mathcal M}(p)$
has been accomplished through the introduction (``by hand'') of a set of functions, 
\mbox{${\mathcal C}_i(q)$}, which in the deep ultraviolet display 
the required asymptotic behavior, but differ substantially at the level of their infrared ``completion''. 
The support of  \mbox{${\mathcal C}_i(q)$}
in the region of \mbox{[$500$ MeV, $1.5$ GeV]} is crucial for the
generation of quark masses of the order of  \mbox{$300$ MeV}. In fact, 
any suppression in the behavior of \mbox{${\mathcal C}_i(q)$}, as reported
in the case  \mbox{${\mathcal C}_3(q)$} given by Eq.~\eqref{rfunc}, can diminish or even eradicate  
the desired phenomenon.
 
The difficulties in enforcing multiplicative renormalizability at the level of the gap equation, as mentioned above, 
make the study of the transverse part of the  quark-gluon vertex
all the more pressing. Even though the relevance of 
$\Gamma^{\Tr}_{\mu}$ in this context has been amply emphasized, 
and various techniques have
been put forth for restricting its structure~\cite{Kondo:1996xn,He:2000we,He:2006my,He:2007zza,Qin:2013mta,Aguilar:2014lha},
a well-defined framework for its systematic determination still
eludes us. In particular, it would be rather important to
obtain reliable results for $\Gamma^{\Tr}_{\mu}$ by means of nonperturbative methods in the continuum
(\eg SDEs~\cite{Roberts:1994dr,Maris:2003vk,Fischer:2006ub,Binosi:2009qm} or functional renormalization group~\cite{Pawlowski:2005xe}), 
especially in view of its theoretical and numerical relevance for chiral symmetry breaking.

As mentioned in the Introduction, 
we have carried out a ``quenched'' calculation, given that the gluon and ghost propagators  used as inputs for solving the system
of integral equations are obtained from lattice simulations with no dynamical quarks~\cite{Bogolubsky:2007ud}. 
To be sure, a more complete analysis ought to take unquenching effects into account; their inclusion  
is expected to affect the results mainly due to the modifications induced to the
gluon propagator (see,~\eg~\cite{Ayala:2012pb} for
unquenched lattice results, and~\cite{Bhagwat:2004kj,Fischer:2005en,Aguilar:2012rz,Williams:2015cvx,Cyrol:2017ewj}
for related studies). A preliminary study presented in~\cite{Aguilar:2014lha}
indicates a slight increase, of the order $6-10\%$, in the form factors of the quark-gluon vertex,
evaluated in some special kinematic limits.
Of course, a complete study needs be carried out
in order to determine if such an increase persists at the level of the coupled system, and the changes that it  might induce to the gap equation and the quark mass derived from it.

\acknowledgments 

The research of J.~P. is supported by the Spanish  Ministerio de Econom\'ia y Competitividad (MEYC) under grants FPA2014-53631-C2-1-P and SEV-2014-0398, and Generalitat Valenciana  
under grant Prometeo~II/2014/066.
The work of  A.~C.~A, J.~C.~C.  and M.~N.~F. are supported by the Brazilian National Council for Scientific and Technological Development (CNPq) under the grants 305815/2015, 141981/2013-0
and  142226/2016-5, respectively. A.~C.~A also acknowledges the financial support
from  S\~{a}o Paulo Research Foundation (FAPESP) through the projects  2017/07595-0 and 
2017/05685-2. This research was performed using the Feynman Cluster of the
John David Rogers Computation Center (CCJDR) in the Institute of Physics ``Gleb
Wataghin", University of Campinas.

%

\end{document}